\documentclass[12pt]{article}
\usepackage[centertags]{amsmath}
\usepackage{amssymb}
\usepackage{rotating}
\usepackage{enumerate}
\usepackage{epsfig}
\usepackage{longtable}
\usepackage{array}
\usepackage{graphics}
\usepackage{graphicx}
\usepackage{color}
\usepackage{bbm}
\usepackage{pst-all}
\usepackage{lscape} 
\usepackage{rotating}
\usepackage[small,loose]{subfigure} 
\usepackage[small]{caption2}

\newcommand{\frF}{\mathfrak{F}}
\newcommand{\frR}{\mathfrak{R}}
\newcommand{\fA}{\mathfrak{U}}
\newcommand{\cA}{\mathcal{A}}
\newcommand{\cF}{\mathcal{F}}
\newcommand{\cR}{\mathcal{R}}
\newcommand{\tr}{\mbox{tr}}
\newcommand{\Tra}{\mbox{Tr}}
\newcommand{\rep}[1]{\boldsymbol{#1}}
\newcommand{\crep}[1]{\overline{\boldsymbol{#1}}}
\newcommand{\as}{&\hspace{-4pt}}
\newcommand{\ass}{&\hspace{-10pt}}
\newcommand{\myspace}{\phantom{$I_{\int}^{\int}$}}
\newcommand{\CP}{\mathbbm{CP}}

\newcommand{\Identity}{{\bf 1}}

\def\fgRelationAxions{
\begin{figure}[t]
\begin{center}
\psset{unit=1.5cm}
\begin{pspicture}(0,0)(7,1)
\psellipse[linewidth=0.4mm](0,0.5)(1,0.7)
\rput(0,0.7){\textbf{\begin{small}heterotic\end{small}}}
\rput(0,0.45){\textbf{\begin{small}orbifold\end{small}}}
\rput(0,0.15){\begin{blue}$\boldsymbol{a^{het}}$\end{blue}}
\psellipse[linewidth=0.4mm](3.5,0.5)(1,0.7)
\rput(3.5,0.7){\textbf{\begin{small}blown-up\end{small}}}
\rput(3.5,0.45){\textbf{\begin{small}orbifold\end{small}}}
\rput(3.5,0.15){\begin{blue}$\boldsymbol{a^{het},\,a^T}$\end{blue}}
\psellipse[linewidth=0.4mm](7,0.5)(1,0.7)
\rput(7,0.7){\textbf{\begin{small}U(1) bundle\end{small}}}
\rput(7,0.45){\textbf{\begin{small}on resolution\end{small}}}
\rput(7,0.15){\begin{blue}$\boldsymbol{a^{mi},\,a^{md}}$\end{blue}}
\psline[linewidth=0.8mm]{->}(1.25,0.5)(2.25,0.5)
\rput(1.75,0.1){\begin{footnotesize}$<\Psi_q>=v e^T$\end{footnotesize}}
\rput(5.25,0.5){\begin{Huge}$\boldsymbol{\cong}$\end{Huge}}
\rput(5.25,0.1){\begin{footnotesize}field red.\end{footnotesize}}
\end{pspicture}
\psset{unit=1cm}
\end{center}
\caption{\label{fig:3} 
Schematic picture of the blowing-up procedure. A twisted field (blow-up mode)
on the orbifold aquires a vev and its phase degree of freedom is reinterpreted as the axion
$a^T$. As U(1)-charges of some of the twisted fields are still different from the ones on the
resolution $\mathcal{M}^3$, appropriate field-redefinitions are necessary to make them
coincide.}
\end{figure}
}

\def\tbVorbClass{
\begin{table}[t]
\begin{center}
\begin{tabular}{|c|c|c|}
\hline&&\\[-2ex]
$V_{orb}$ & gauge group & \\[1ex]
\hline\hline&&\\[-2ex]
$(0^8)(0^8)$ & E$_8\times$E$_8'$ &\\[.5ex]
$(-2,1^2,0^5)(0^8)$ & E$_6\times$SU(3)$\times$E$_8'$ &A\\[.5ex]
$(-2,1^2,0^5)(-2,1^2,0^5)$ &$\big[$E$_6\times$SU(3)$\big]^2$ &B\\[.5ex]
$(1^2,0^6)(-2,0^7)$ & E$_7\times$SO(14)$\times$U(1)$^2$ &C\\[.5ex]
$(-2,1^4,0^3)(-2,0^7)$ & SU(9)$\times$SO(14)$\times$U(1) &D\\[.5ex]
\hline
\end{tabular}
\caption{\label{orbshift}
We list all inequivalent $\mathbbm C^3/\mathbbm Z_3$ orbifold shifts and the 
corresponding gauge symmetry breakings. For shifts, expressions like $0^n$ denote 
$n$ times the zero entry.
}
\end{center}
\end{table}
}

\def\tbVblowClass{
\begin{table}[t]
\begin{center}
{\footnotesize
\begin{tabular}{|ccc|c|c|}
\hline
 & bundle vector  & & gauge group & label \\
 & $V = (V_1)(V_2)$  & & $V_1^2 + V_2^2 = V^2 = 12$       &   \phantom{$I^{I^I}$}     \\
\hline
$\left(3,1^3,0^4\right)\left(0^8\right)$                                   & $\left(2^3,0^5\right)\left(0^8\right)$                                          & $\left(2^2,1^4,0^2\right)\left(0^8\right)$                                         & $\text{SO}(10)\times \text{U}(3) \times \text{E}'_8$      & \phantom{$I^{I^I}$}\hspace{-0.5cm}AI \\
$\left(\frac{5}{2}, \frac{3}{2}^2,\frac{1}{2}^5\right)\left(0^8\right)$    & $\left(\frac{3}{2}^4,-\frac{3}{2},\frac{1}{2}^3\right)\left(0^8\right)$         &                                                                                    & $12 + 0$                                                  &    \\
\hline
$\left(2,1^2,0^5\right)\left(2,1^2,0^5\right)$                             & $\left(2,1^2,0^5\right)\left(1^6,0^2\right)$                                    & $\left(2,1^2,0^5\right)\left(\frac{3}{2}^2,\frac{1}{2}^6\right)$                   & $\left(\text{E}_6 \times \text{U}(2)\right)^2$            & \phantom{$I^{I^I}$}\hspace{-0.5cm}BI \\
$\left(1^6,0^2\right)\left(1^6,0^2\right)$                                 & $\left(1^6,0^2\right)\left(\frac{3}{2}^2,\frac{1}{2}^6\right)$                  & $\left(\frac{3}{2}^2,\frac{1}{2}^6\right)\left(\frac{3}{2}^2,\frac{1}{2}^6\right)$ & $6 + 6$                                                   &    \\
\hline
$\left(2^2,0^6\right)\left(2,0^7\right)$                                   & $\left(2^2,0^6\right)\left(1^4,0^4\right)$                                      & $\left(2^2,0^6\right)\left(-\frac{3}{2},\frac{1}{2}^7\right)$                      & $\text{E}_7 \times \text{SO}(14)' \times \text{U}(1)^2$   & \phantom{$I^{I^I}$}\hspace{-0.5cm}CI \\
$\left(1^8\right)\left(2,0^7\right)$                                       & $\left(1^8\right)\left(1^4,0^4\right)$                                          & $\left(1^8\right)\left(-\frac{3}{2},\frac{1}{2}^7\right)$                          & $8 + 4$                                                   &    \\
\hline
$\left(1^2, 0^6\right)\left(3,1,0^6\right)$                                & $\left(1^2, 0^6\right)\left(2^2,1^2,0^4\right)$                                 & $\left(1^2, 0^6\right)\left(2,1^6,0\right)$                                        & $\text{E}_7 \times \text{SO}(12)' \times \text{U}(1)^3$   & \phantom{$I^{I^I}$}\hspace{-0.5cm}CII \\
$\left(1^2, 0^6\right)\left(\frac{5}{2},-\frac{3}{2},\frac{1}{2}^6\right)$ & $\left(1^2,0^6\right)\left(\frac{3}{2}^4,\frac{1}{2}^4\right)$                  & $\left(\frac{1}{2}^8\right)\left(3,1,0^6\right)$                                   & $2 + 10$                                                  &     \\

$\left(\frac{1}{2}^8\right)\left(2^2,1^2,0^4\right)$                       & $\left(\frac{1}{2}^8\right)\left(2,1^6,0\right)$                                & $\left(\frac{1}{2}^8\right)\left(\frac{5}{2},-\frac{3}{2},\frac{1}{2}^6\right)$    &                                                           &     \\
$\left(\frac{1}{2}^8\right)\left(\frac{3}{2}^4,\frac{1}{2}^4\right)$       &                                                                                 &                                                                                    &                                                           &     \\
\hline
$\left(2,1^4,0^3\right)\left(2,0^7\right)$                                 & $\left(2,1^4,0^3\right)\left(1^4,0^4\right)$                                    & $\left(2,1^4,0^3\right)\left(-\frac{3}{2},\frac{1}{2}^7\right)$                    & $\text{SU}(8) \times \text{SO}(14)' \times \text{U}(1)^2$ & \phantom{$I^{I^I}$}\hspace{-0.5cm}DI \\
$\left(-\frac{3}{2}^3,\frac{1}{2}^5\right)\left(2,0^7\right)$              & $\left(-\frac{3}{2}^3,\frac{1}{2}^5\right)\left(1^4,0^4\right)$                 & $\left(-\frac{3}{2}^3,\frac{1}{2}^5\right)\left(-\frac{3}{2},\frac{1}{2}^7\right)$ & $8 + 4$                                                   &    \\
$\left(\frac{5}{2},\frac{1}{2}^7\right)\left(2,0^7\right)$                 & $\left(\frac{5}{2},\frac{1}{2}^7\right)\left(1^4,0^4\right)$                    & $\left(\frac{5}{2},\frac{1}{2}^7\right)\left(-\frac{3}{2},\frac{1}{2}^7\right)$    &                                                           &    \\
$\left(-1,1^7\right)\left(2,0^7\right)$                                    & $\left(-1,1^7\right)\left(1^4,0^4\right)$                                       & $\left(-1,1^7\right)\left(-\frac{3}{2},\frac{1}{2}^7\right)$                       &                                                           &    \\
\hline
\end{tabular} 
}
\caption{\label{tab:E8shifts}
In this table we list all consistent U(1) bundles embedded into 
$\text{E}_8 \times \text{E}'_8$. We group together the embeddings
characterized by the bundle vector $V$ producing the same 
gauge symmetry breaking and localized spectrum.
Each group corresponds to a distinct blow--up of the orbifold
models. The bundle vector $V$ contains two parts corresponding to
both E$_8$'s. Most models are characterized by the values of $V_1^2$ and 
$V_2^2$; only the splitting $8+4$ has two realizations.}
\end{center}
\end{table}
}

\def\tbMatchingZthree{
\begin{landscape}
\begin{table}[t]
\begin{center}
\begin{tabular}{|c||c|c|c|}
\hline&&&\\[-2ex]
 & orbifold model & resolution model & field redefinitions\\[.5ex]
\hline\hline&&&\\[-2ex]
& $\text{E}_6\times \text{SU(3)} \times \text{E}_8'$ & 
$\text{SO(10)}\times \text{U(3)} \times \text{E}_8'$ &
\\[.5ex]\cline{2-3}&&&\\[-2ex]
\hspace{-10pt} 
\begin{tabular}{c} 
A \\ $\downarrow$ \\ AI 
\end{tabular} 
\hspace{-11pt}
&
$\begin{array}{c}
\frac{1}{9} (\rep{27},\rep{3};\rep{1})\\+
{(\rep{27},\rep{1};\rep{1})}+
3(\rep{1},\crep{3};\rep{1})
\end{array}$
&\mbox{}\hspace{-12pt}
$\begin{array}{c}
\frac{1}{9}\left[ (\rep{16},\rep{3};\rep{1})_{\text{-}1}+
(\rep{10},\rep{3};\rep{1})_{2}+
(\rep{1},\rep{3};\rep{1})_{\text{-}4}\right]\\
+(\rep{16},\rep{1};\rep{1})_{3}
+3 (\rep{1},\crep{3};\rep{1})_{4}
\end{array}$\hspace{-8pt}
&
$\begin{array}{lll}\vspace{-36pt}
\as\as\\
(\rep{27},\rep{1};\rep{1})
\as  \hspace{-5pt} \rightarrow \as 
\hspace{-5pt}
\left\{ \hspace{-6pt} 
\begin{array}{lcl}
(\rep{1},\rep{1};\rep{1})_{\text{-}4}\ass=\as e^T v\\
(\rep{16},\rep{1};\rep{1})_{\text{-}1}\ass=\as e^T (\rep{16},\rep{1};\rep{1})_{3}\\
(\rep{10},\rep{1};\rep{1})_{2}\ass=\as e^{-T/2} (\rep{10},\rep{1};\rep{1})^{m}_{0}
\end{array}\right. 
\\[4ex] 
(\rep{1},\crep{3};\rep{1})
\as  \hspace{-5pt} \rightarrow \as  
\hspace{6pt} (\rep{1},\crep{3};\rep{1})_{0} = e^T
(\rep{1},\crep{3};\rep{1})_{4}
\end{array} 
$\\[2.5ex]
\hline\hline&&&\\[-2ex]
& 
$\left[\text{E}_6\times \text{SU(3)}\right]^2$ & 
$\left[\text{E}_6\times \text{U(2)}\right]^2$ &
\\[.5ex]\cline{2-3}&&&\\[-2ex]
\hspace{-10pt}
\begin{tabular}{c} 
B \\ $\downarrow$ \\ BI 
\end{tabular} 
\hspace{-11pt}
&
$\begin{array}{c}
\frac{1}{9} \left[(\rep{27},\crep{3};\rep{1},\rep{1})\right.\\\left.+
(\rep{1},\rep{1};\rep{27},\crep{3})
\right]\\
+{(\rep{1},\rep{3};\rep{1},\rep{3})}
\end{array}$
&
$\begin{array}{c}
\frac{1}{9}\left[ (\rep{27},\rep{2};\rep{1},\rep{1})_{\text{-}1,\text{-}1}+
 (\rep{27},\rep{1};\rep{1},\rep{1})_{2,2}\right.\\
\left.+(\rep{1},\rep{1};\rep{27},\rep{2})_{\text{-}1,1}+
 (\rep{1},\rep{1};\rep{27},\rep{1})_{2,\text{-}2}
\right]\\
+(\rep{1},\rep{2};\rep{1},\rep{1})_{3,3}+
(\rep{1},\rep{1};\rep{1},\rep{2})_{3,\text{-}3}
\end{array}$
&\mbox{}\hspace{-12pt}
$\begin{array}{lll}
\vspace{-36pt}\as\as\\
(\rep{1},\rep{3};\rep{1},\rep{3})
\as  \hspace{-5pt} \rightarrow \as 
\hspace{-5pt}
\left\{ \hspace{-6pt} 
\begin{array}{lcl}
(\rep{1},\rep{1};\rep{1},\rep{1})_{\text{-}4,0}\ass=\as e^T v\\
(\rep{1},\rep{2};\rep{1},\rep{1})_{\text{-}1,3}\ass=
\as e^T (\rep{1},\rep{2};\rep{1},\rep{1})_{3,3}\\
(\rep{1},\rep{1};\rep{1},\rep{2})_{\text{-}1,\text{-}3}\ass=
\as e^T (\rep{1},\rep{1};\rep{1},\rep{2})_{3,\text{-}3}\\
(\rep{1},\rep{2};\rep{1},\rep{2})_{2,0}\ass=
\as e^{-T/2} (\rep{1},\rep{2};\rep{1},\rep{2})^{m}_{0,0}
\end{array}
\right. 
\end{array} 
$\hspace{-14pt}
\\[3ex]
\hline\hline&&&\\[-2ex]
& 
$\text{E}_7\times \text{SO(14)}\times \text{U(1)}^2$ & 
$\text{E}_7\times \text{SO(14)}\times \text{U(1)}^2$ &
\\[.5ex]\cline{2-3}&&&\\[-2ex]
\hspace{-10pt}
\begin{tabular}{c} 
C \\ $\downarrow$ \\ CI 
\end{tabular} 
\hspace{-11pt}
&
\mbox{}\hspace{-12pt}
$\begin{array}{c}
\frac{1}{9} \left[(\rep{56};\rep{1})_{2,2}
+(\rep{1};\rep{1})_{\text{-}4,\text{-}4} \right.\\\left.
+(\rep{1};\rep{64})_{\text{-}1,2}
+(\rep{1};\rep{14})_{2,\text{-}4}
\right]+ \\
(\rep{1};\rep{14})_{2,0}\hspace{-4pt}
+{(\rep{1};\rep{1})}_{\text{-}4,0}\hspace{-4pt}
+3 (\rep{1};\rep{1})_{0,4}
\end{array}$\hspace{-8pt}
&
$\begin{array}{c}
\frac{1}{9} \left[(\rep{56};\rep{1})_{2,2}
+(\rep{1};\rep{1})_{\text{-}4,\text{-}4} \right.\\\left.
+(\rep{1};\rep{64})_{\text{-}1,2}
+(\rep{1};\rep{14})_{2,\text{-}4}
\right]\\
+3 (\rep{1};\rep{1})_{4,4}
\end{array}$
&
$\begin{array}{lcl}
\vspace{-36pt}\as\as\\
(\rep{1};\rep{1})_{\text{-}4,0} \ass = \as e^T v
\\[1ex] 
 (\rep{1};\rep{1})_{0,4}
 \ass = \as e^{T} (\rep{1};\rep{1})_{4,4} 
\\[1ex] 
(\rep{1};\rep{14})_{2,0}
\ass = \as e^{-T/2}(\rep{1};\rep{14})_{0,0}^{m}
\end{array}$
\\[3ex]
\hline\hline&&&\\[-2ex]
& 
$\text{E}_7\times \text{SO(14)}\times \text{U(1)}^2$ & 
$\text{E}_7\times \text{SO(12)}\times \text{U(1)}^3$ &
\\[.5ex]\cline{2-3}&&&\\[-2ex]
\hspace{-10pt}
\begin{tabular}{c} 
C \\ $\downarrow$ \\ CII 
\end{tabular} 
\hspace{-11pt}
&
\mbox{}\hspace{-12pt}
$\begin{array}{c}
\frac{1}{9} \left[(\rep{56};\rep{1})_{2,2}
+(\rep{1};\rep{64})_{\text{-}1,2} \right.\\\left.
+(\rep{1};\rep{1})_{\text{-}4,\text{-}4} 
+(\rep{1};\rep{14})_{2,\text{-}4}
\right]+\\
{(\rep{1};\rep{14})}_{2,0}\hspace{-4pt}
+(\rep{1};\rep{1})_{\text{-}4,0}\hspace{-4pt}
+3 (\rep{1};\rep{1})_{0,4}
\end{array}$\hspace{-8pt}
&
\mbox{}\hspace{-12pt}
$\begin{array}{c}
\frac{1}{9} [(\rep{56};\rep{1})_{\text{-}1,2
,\text{-}2}\hspace{-3pt}
+(\rep{1};\rep{32})_{\text{-}1,2
,2}+
(\rep{1};\crep{32})_{2,2
,0}
\\
+(\rep{1};\rep{1})_{\text{-}4,\text{-}4
,0}
\!+\! (\rep{1};\rep{12})_{\text{-}1,\text{-}4
,\text{-}2} \!+
(\rep{1};\rep{1})_{2,\text{-}4
,\pm 4}]+
\\ 
(\rep{1};\rep{12})_{3,0,\text{-}2}
\!+\! 3 (\rep{1};\rep{1})_{4,4
,0}
\end{array}$\hspace{-8pt}
&
$
\begin{array}{lll}
\vspace{-40pt}\as\as\\
(\rep{1};\rep{14})_{2,0}
\as  \hspace{-5pt} \rightarrow \as 
\hspace{-5pt}
\left\{ \hspace{-6pt} 
\begin{array}{lcl}
(\rep{1};\rep{1})_{\text{-}4,0,0}\ass=\as e^T v\\ 
(\rep{1};\rep{12})_{\text{-}1,0,\text{-}2}\ass=
\as e^T (\rep{1};\rep{12})_{3,0,\text{-}2}\\
(\rep{1};\rep{1})_{2,0,\text{-} 4}\ass=
\as e^{-T/2} (\rep{1};\rep{1})^{m}_{0,0,\text{-}4}
\end{array}
\right. 
\\[4ex]  
(\rep{1};\rep{1})_{\text{-}4,0}\hspace{-4pt}
\as  \hspace{-5pt} \rightarrow \as 
\begin{array}{lcl}
(\rep{1};\rep{1})_{2,0,4}\ass=
\as e^{-T/2} (\rep{1};\rep{1})^{m}_{0,0,4}
\end{array}
\\[1ex]  
(\rep{1};\rep{1})_{0,4}
\as \hspace{-5pt} \rightarrow \as 
\begin{array}{lcl}
(\rep{1};\rep{1})_{0,4
,0}\ass=
\as e^T (\rep{1};\rep{1})_{4,4
,0}\\
\end{array}
\end{array}
$
\\\hline\hline&&&\\[-2ex]
&
$\text{SU(9)}\times \text{SO(14)}\times \text{U(1)}$ & 
$\text{SU(8)}\times \text{SO(14)}\times \text{U(1)}^2$ &
\\[.5ex]\cline{2-3}&&&\\[-2ex]
\hspace{-10pt}
\begin{tabular}{c} 
D \\ $\downarrow$ \\ DI 
\end{tabular}
\hspace{-11pt}  
&
\mbox{}\hspace{-12pt}
$\begin{array}{c}
\frac{1}{9} \left[(\rep{84};\rep{1})_{0}\hspace{-4pt}
+(\rep{1};\rep{64})_{\text{-}1}\hspace{-4pt}
+(\rep{1};\rep{14})_{2}
\right]\\
+{(\crep{9};\rep{1})}_{\text{-}4/3}
\end{array}$\hspace{-8pt}
&
$\begin{array}{c}
\frac{1}{9} \left[(\rep{56};\rep{1})_{\text{-}1,\text{-}1}\hspace{-4pt}+
(\rep{28};\rep{1})_{2,2}\hspace{-4pt}
+(\rep{1};\rep{64})_{\text{-}1,2}\right.\\\left.
+(\rep{1};\rep{14})_{2,\text{-}4}
\right]
+(\crep{8};\rep{1})_{3,3}
\end{array}$\hspace{-8pt}
&
$ \begin{array}{l}
\vspace{-36pt} \\
(\crep{9};\rep{1})_{\text{-}4/3} \rightarrow 
\left\{ \begin{array}{l}
(\rep{1};\rep{1})_{\text{-}4,0}  =  e^T v
\\[1ex] 
 (\crep{8};\rep{1})_{\text{-}1,3}
 =  e^{T} (\crep{8};\rep{1})_{3,3} 
\end{array}
\right. 
\end{array}
$
\\[2ex]\hline
\end{tabular} 
\caption{\label{tab:match}
We define the matching orbifold and blow--up models in the first
column. The second and third columns give their orbifold and resolution spectra,
respectively. The final column gives the field redefinitions necessary
to match the two spectra.  For blow--up models CII and DI a change of
U(1) basis accompanies the branching (indicated by $\rightarrow$) to
ensure that the state getting the vev is charged under the first blow--up
U(1) only.  
The superscript $m$ indicates non--chiral states that get a mass in
blow--up, and therefore decouple from the massless spectrum.}
\end{center}
\end{table}
\end{landscape}
}

\def\tbAnomPolyZthree{ 
\begin{table}[t]\footnotesize
\begin{center}
\begin{tabular}{|c||l|c|}
\hline&\multicolumn{1}{|c|}{}&\\[-2ex]
&
\multicolumn{1}{|c|}{anomaly polynomials} &
$\alpha$ \\
\hline\hline&&\\[-2ex]
\mbox{}\hspace{-12pt}
\begin{tabular}{c}A \\ $\downarrow$\\ AI\end{tabular}
\hspace{-10pt}
&
$
\begin{array}{l}
\hat I_6^{het}=0\\[1ex]
\hat I_6^{uni} \,= 
\frac{3}{2}(i F)\left[24 (i F)^2 + (i F_{10})^2 +
 2(i F_3)^2 +(i F_8)^2 -  R^2 \right]\\[1ex]
\hat I_6^{non} = \frac{1}{4}(i F)\left[528 (i F)^2 + 
6 (i F_{10})^2 + 12(i F_3)^2 - 6(i F_8)^2 -R^2 \right]
\end{array}$
& $-\frac{3}{8}$
\\[5ex]\hline&&\\[-2ex]
\mbox{}\hspace{-12pt}
\begin{tabular}{c}B \\ $\downarrow$\\ BI\end{tabular}
\hspace{-10pt}
&
$\begin{array}{ll}
\hat I_6^{het}=0\\[1ex]
\hat I_6^{uni} \,= 0\\[1ex]
\hat I_6^{non} = \frac 14 (i F)\left[96 (i F)^2 + 
288 (i F')^2  -R^2 \right]\\[1ex]
\end{array}$
& $0$
\\[1ex]\hline&&\\[-2ex]
\mbox{}\hspace{-12pt}
\begin{tabular}{c}C \\ $\downarrow$\\ CI\end{tabular}
\hspace{-10pt}
& 
$\begin{array}{l}
\hat I_6^{het}  = \frac{2}{3} (i F +i F')
\left[24 (i F)^2 + 48 (i F')^2 + \frac{1}{6} (i F_7)^2 +
(i F_{14})^2 - R^2\right]\\[1ex] 
\hat I_6^{uni} \,= \frac{1}{6} (i F +4 i F')\left[24 (i F)^2 + 48 (i F')^2 + 
\frac{1}{6} (i F_7)^2 +(i F_{14})^2 - R^2\right]
\\[1ex]
\hat I_6^{non} = \frac{1}{4}(i F)
\left[144 (i F)^2 + 480 (i F')^2+384 (i F)(i F')
+\frac{1}{3} (i F_7)^2 - 2(i F_{14})^2 - R^2\right]
\end{array}$
& $\frac{1}{8}$
\\[5ex]\hline&&\\[-2ex]
\mbox{}\hspace{-12pt}
\begin{tabular}{c}C \\ $\downarrow$\\ CII\end{tabular}
\hspace{-10pt}
&
$\begin{array}{l}
\hat I_6^{het}  =   -\frac{1}{3}( i F - \frac 12 i F' +2 i F'')\times\\[.5ex]
\mbox{}\hspace{15pt}
\times \left[24 (i F)^2 + 3 (i F')^2 + 32 (i F'')^2+
 \frac{1}{6} (i F_7)^2 + (i F_{12})^2 - R^2\right]
\\[1ex]
\hat I_6^{uni} \,= 
\frac{1}{12}(17 i F +2 i F' -8 i F'')\times\\[.5ex]
\mbox{}\hspace{15pt}\times
\left[24 (i F)^2 + 3 (i F')^2 + 32 (i F'')^2 
+ \frac{1}{6} (i F_7)^2 + (i F_{12})^2 - R^2\right]
\\[1ex]
\hat I_6^{non} = \frac{1}{4}(i F)
\left[288 (i F)^2 + 12 (i F')^2 + 128 (i F'')^2 
+96 (i F)(i F') \right.\\[.5ex]\left.\mbox{}\hspace{24pt}
- 384 (i F)(i F'') 
-\frac{2}{3} (i F_7)^2 + 4 (i F_{12})^2 - R^2 \right]
\end{array}$
& $-\frac{1}{4}$
\\[10ex]\hline&&\\[-2ex]
\mbox{}\hspace{-12pt}
\begin{tabular}{c}D \\ $\downarrow$\\ DI\end{tabular}\hspace{-10pt}
&
$\begin{array}{l}
\hat I_6^{het} = 
-\frac{1}{3}(i F-2iF') \left[24 (i F)^2 +48 (i F')^2+ 2(i F_8)^2 +(i F_{14})^2 - R^2\right]\\[1ex]
\hat I_6^{uni} \,= \frac 16 (i F +4 i F')
\left[24 (i F)^2 + 48 (i F')^2 + 2 (i F_8)^2 + (i F_{14})^2 -
R^2\right]
\\[1ex]
\hat I_6^{non} =\frac 14 (i F)
\left[192 (i F)^2 + 480 (i F')^2 +384 (i F)(i F')
+ 4 (i F_8)^2 - 2(i F_{14})^2 - R^2\right]
\end{array}$
& $-\frac{1}{8}$
\\[5ex]\hline
\end{tabular} 
\caption{\label{tab:anomal}
The anomalies of the blow-ups are compared with those of the
orbifold theories. The resolution anomaly polynomial $\hat I^{blow}_6$
is divided into a universal part $\hat I_6^{uni}$ and a non-universal
part $\hat I_6^{non}$. The axion redefinition parameter $\alpha$ is
defined in Eq.~\eqref{eq:axions}. Note that we omitted the trace tr
for the curvature and all non-abelian gauge group factors.  
}
\end{center}
\end{table}
}

\def\fgCompactBlowupVEVs{
\begin{table}[t]\footnotesize
\begin{center}
\begin{tabular}{|c||c|l|l|ll|}
\hline
fixed point      & 
\multicolumn{3}{c|}{matter}      & 
decomposition    & 
\mbox{}\hspace{-6pt}field  \\
\cline{2-4}
(loc.) gauge group & 
\mbox{}\hspace{-4pt}mult.\hspace{-6pt} &
local matt. & 
4D matt. & 
to blow-up group & 
\mbox{}\hspace{-6pt}redefinition \\
\hline\hline&&&&&\\[-2ex]
U Sector   & 3 &  & 
$(\boldsymbol{27}, \boldsymbol{1})_{(2,2,0)}$ & 
$(\boldsymbol{16}, \boldsymbol{1})_{(2,2,0,-1)}$ &
\\[.5ex]
$\text{E}_6\times\text{SO}(14)\times\text{U}(1)^3$ && & &
$(\boldsymbol{10}, \boldsymbol{1})_{(2,2,0,2)}$ &
\\[.5ex] &   &&&
$(\boldsymbol{1}, \boldsymbol{1})_{(2,2,0,-4)}$  &
\\[.5ex]
\hline\hline&&&&&\\[-2ex]
$g_1 = \left( \theta, 0\right)$ & 
1 & 
$(\boldsymbol{1}, \boldsymbol{14})_{(2,0)}$ & 
$(\boldsymbol{1}, \boldsymbol{14})_{(2,0,0)}$ & 
$(\boldsymbol{1}, \boldsymbol{14})_{(2,0,0,0)}$
& \mbox{}\hspace{-30pt}
$=e^{-\frac{1}{2}T_1}(\boldsymbol{1}, \boldsymbol{14})^m_{(0,0,0,0)}$\\[.5ex]
\cline{2-6}&&&&&\\[-2ex]
$\text{E}_7\times\text{SO}(14)\times\text{U}(1)^2$ & 
1 & 
$(\boldsymbol{1}, \boldsymbol{1})_{(-4,0)}$ & 
$(\boldsymbol{1}, \boldsymbol{1})_{(-4,0,0)}$ & 
$(\boldsymbol{1}, \boldsymbol{1})_{(-4,0,0,0)}$
&\mbox{}\hspace{-30pt}
$=v_1 e^{T_1}$\\[.5ex]
\cline{2-6}&&&&&\\[-2ex]
& 3 & 
$(\boldsymbol{1}, \boldsymbol{1})_{(0,4)}$  & 
$(\boldsymbol{1}, \boldsymbol{1})_{(0,2,-2)}$ & 
$(\boldsymbol{1}, \boldsymbol{1})_{(0,2,-2,0)}$
& \mbox{}\hspace{-30pt}
$=e^{T_1}(\boldsymbol{1}, \boldsymbol{1})_{(4,2,-2,0)}$\\[.5ex]
\cline{2-6}
local blow-up at $g_1$ & 
\multicolumn{5}{|l|}{CI\myspace}\\
\hline\hline
$g_2 = \left( \theta, e_1\right)$ & 1 & 
$(\boldsymbol{27}, \boldsymbol{1}, \boldsymbol{1})$ &
$(\boldsymbol{27}, \boldsymbol{1})_{(0,0,0)}$  & 
$(\boldsymbol{16}, \boldsymbol{1})_{(0,0,0,-1)}$
\myspace & \mbox{}\hspace{-30pt}
$=e^{T_2}(\boldsymbol{16}, \boldsymbol{1})_{(0,0,0,3)}$\\
$\text{E}_6\times\text{SU}(3)\times\text{E}_8$     &&&&
$(\boldsymbol{10}, \boldsymbol{1})_{(0,0,0,2)}$
\myspace  & \mbox{}\hspace{-30pt}
$=e^{-\frac{1}{2}T_2}( \boldsymbol{10}, \boldsymbol{1})^m_{(0,0,0,0)}$
\\ &&&& 
$(\boldsymbol{1}, \boldsymbol{1})_{(0,0,0,-4)}$
\myspace  & \mbox{}\hspace{-30pt}
$=v_2 e^{T_2}$\\
\cline{2-6} & 3 &
$(\boldsymbol{1}, \boldsymbol{3}, \boldsymbol{1})$ &
$(\boldsymbol{1}, \boldsymbol{1})_{(-2,-2,0)}$ &
$(\boldsymbol{1}, \boldsymbol{1})_{(-2,-2,0,0)}$
\myspace & \mbox{}\hspace{-30pt}
$=e^{T_2}(\boldsymbol{1}, \boldsymbol{1})_{(-2,-2,0,4)}$
\\ &&&
$(\boldsymbol{1}, \boldsymbol{1})_{(0,2,2)}$   & 
$(\boldsymbol{1}, \boldsymbol{1})_{(0,2,2,0)}$
\myspace   & \mbox{}\hspace{-30pt}
$=e^{T_2}(\boldsymbol{1}, \boldsymbol{1})_{(0,2,2,4)}$
\\ &&&
$(\boldsymbol{1}, \boldsymbol{1})_{(2,0,-2)}$  & 
$(\boldsymbol{1}, \boldsymbol{1})_{(2,0,-2,0)}$
\myspace  &\mbox{}\hspace{-30pt}
$=e^{T_2}(\boldsymbol{1}, \boldsymbol{1})_{(2,0,-2,4)}$\\
\cline{2-6}
local blow-up at $g_2$ & \multicolumn{5}{|l|}{AI\myspace}\\
\hline\hline
$g_3 = \left( \theta, e_1 + e_2\right)$& 1 & 
$(\boldsymbol{1}, \boldsymbol{14})_{(0,2)}$ & 
$(\boldsymbol{1}, \boldsymbol{14})_{(0,2,0)}$ & 
$(\boldsymbol{1}, \boldsymbol{14})_{(0,2,0,0)}$
\myspace & \mbox{}\hspace{-30pt}
$=e^{-\frac{1}{2}T_3}(\boldsymbol{1}, \boldsymbol{14})^m_{(0,0,0,0)}$\\
\cline{2-6}
$\text{E}_7\times\text{SO}(14)\times\text{U}(1)^2$ & 1 & 
$(\boldsymbol{1}, \boldsymbol{1})_{(0,-4)}$ & 
$(\boldsymbol{1}, \boldsymbol{1})_{(0,-4,0)}$ & 
$(\boldsymbol{1}, \boldsymbol{1})_{(0,-4,0,0)}$
\myspace & \mbox{}\hspace{-30pt}
$=v_3 e^{T_3}$\\
\cline{2-6} & 3 & 
$(\boldsymbol{1}, \boldsymbol{1})_{(4,0)}$  & 
$(\boldsymbol{1}, \boldsymbol{1})_{(2,0,2)}$  & 
$(\boldsymbol{1}, \boldsymbol{1})_{(2,0,2,0)}$
\myspace  & \mbox{}\hspace{-30pt}
$=e^{T_3}(\boldsymbol{1}, \boldsymbol{1})_{(2,4,2,0)}$\\
\cline{2-6} 
local blow-up at $g_3$ & \multicolumn{5}{|l|}{CI\myspace}\\
\hline
\end{tabular}
\caption{\label{tab:table2}
This table gives an overview of the complete global 4D spectrum of the
blown up orbifold theory. The field redefinitions necessary to have
precisely local matching between the orbifold blow--up theory and the
resolution model are indicated. The $\text{U}(1)^4$-generators of the
4D gauge group in blow--up are $Q_1 = (2,2,0^6)(2,0^7)$, $Q_2 =
(2,0,-2,0^5)(-2,0^7)$, $Q_3 = (0,-2,-2,0^5)(2,0^7)$ and
 $Q_4 = (2,-2,2,0^5)(0^8)$. There are two anomalous combinations: 
$Q_1^{an} = Q_1+Q_2$ and $Q_2^{an} = Q_4$. }
\end{center}
\end{table}
}

\def\tbVevsZthreeMSSM{ 
\begin{table}[t]
\begin{center}
\begin{tabular}{|c|rr|rrrrrrrr|c|c|}
\hline
state & \multicolumn{2}{c}{fixed point} & \multicolumn{8}{|c|}{U(1) charges}                            & \multicolumn{1}{c|}{hyper} & local\\
label & $~~~n_1$ & $n_3$                & $Q_1$ & $Q_2$ & $Q_3$ & $Q_4$ & $Q_5$ & $Q_6$ & $Q_7$ & $Q_8$ &  charge $Y$                & blow-up\\
\hline
\hline
$h_2$    & 0 & 0 & -3 & -2 &  3 &  3 & -3 &  4 &  0 &  0 &   0 & DI\\ 
$h_{10}$ & 1 & 0 & -3 & -2 &  3 &  3 &  1 & -2 &  2 & -4 &   0 & BI\\ 
$h_{14}$ &-1 & 0 &  6 &  4 &  0 &  0 &  2 &  4 & -2 & -2 &   0 & BI\\ 
$h_{15}$ & 0 & 1 & -6 &  0 &  0 &  2 & -4 &  0 & -4 &  0 &   0 & DI\\ 
$h_{17}$ & 0 &-1 &  0 & -4 &  0 & -2 & -2 & -4 &  4 &  0 &   0 & CI\\ 
$h_{21}$ & 1 & 1 & -6 &  0 &  0 &  2 &  0 &  0 &  4 & -4 &   0 & CI\\ 
$h_{23}$ &-1 & 1 &  3 &  6 & -3 & -1 &  1 &  0 &  0 &  4 &   0 & DI\\ 
$h_{24}$ & 1 &-1 &  0 & -4 &  0 & -2 &  2 & -4 &  0 & -4 &   0 & CI\\ 
\hline
\end{tabular}
\end{center}
\caption{\label{tab:Z3-2WLblowupmodes}
The eight blow-up modes --one per resolved fixed point-- are 
chosen to be singlets with respect to 
$\text{SU}(3)\times\text{SU}(2)_L\times\text{U}(1)_Y$. 
The notation used here follows~\cite{Casas:1988se}.}
\end{table}
}

\def\tbMatchingBlowupZtwo{ 
\begin{table}[t]\footnotesize
\begin{center}
\begin{tabular}{|c||c|c|c|}
\hline&&&\\[-2ex]
matching & orbifold model & resolution model & field redefinitions\\[.5ex]
\hline\hline&&&\\[-2ex]
2A $\rightarrow$ 2AI & $\text{SO(28)}\times \text{SU(2)}^2$ & 
$\text{SO(26)}\times \text{SU(2)} \times \text{U(1)}^2$ &
\\[.5ex]\cline{2-3}&&&\\[-2ex]
$\begin{array}{c}\langle (\rep{28},\rep{1},\rep{2})\rangle \\ \neq 0
\end{array}$
&\mbox{}\hspace{-12pt}
$\begin{array}{c}
\frac{1}{16} (\rep{28},\rep{2},\rep{2})+\\
\frac 12 (\rep{28},\rep{1},\rep{2})+
2(\rep{1},\rep{2};\rep{1})
\end{array}$\hspace{-10pt}
&\mbox{}\hspace{-12pt}
$\begin{array}{c}
\frac{1}{8}\left[ (\rep{26},\rep{2})_{1,1}+
(\rep{1},\rep{2})_{2\pm 1,-1\pm 1}\right]\\
+\frac{7}{8}\left[(\rep{26},\rep{1})_{2,\text{-}1}+(\rep{1},\rep{1})_{2,2}\right]\\
+2 (\rep{1},\rep{2})_{3,0}
\end{array}$\hspace{-8pt}
&\mbox{}\hspace{-12pt}
$\begin{array}{lcl}\vspace{-36pt}
\ass\ass\\
(\rep{1},\rep{1})_{3,0}\ass=\as e^T v\\
(\rep{26},\rep{1})_{\text{-}1,\text{-}1}\ass=\as e^{-T} (\rep{26},\rep{1})_{2,\text{-}1}\\
(\rep{1},\rep{2})_{0,0}\ass=\as e^{-T} (\rep{1},\rep{2})_{3,0}\\
(\rep{1},\rep{1})_{\text{-}1,2}\ass=\as e^{-T} (\rep{1},\rep{1})_{2,2}
\end{array}$\hspace{-8pt}
\\[2.5ex]
\hline\hline&&&\\[-2ex]
2B $\rightarrow$ 2BI & $\text{SO(20)}\times \text{SO(12)}$ &
$\text{SO(20)}\times \text{U(6)}$ &
\\[.5ex]\cline{2-3}&&&\\[-2ex]
$\langle(\rep{1},\rep{32})\rangle \neq 0$
&\mbox{}\hspace{-12pt}
$\begin{array}{c}
\frac{1}{16} (\rep{20},\rep{12}) +
\frac{1}{2}(\rep{1},\rep{32})
\end{array}$\hspace{-8pt}
&
$\begin{array}{c}
\frac{1}{8} (\rep{20},\rep{6})_1 +
\frac{7}{8}(\rep{1},\rep{15})_2
\end{array}$
&
$\begin{array}{lcl}
\vspace{-36pt}
\ass\ass\\
(\rep{1},\rep{1})_{3}\ass=\as e^T v\\
(\rep{1},\rep{15})_{\text{-}1}\ass=\as e^{-T} (\rep{1},\rep{15})_{2}
\end{array}$
\\[1ex]
\hline\hline&&&\\[-2ex]
2C $\rightarrow$ 2CI & U(16) &
$\text{SU(15)}\times \text{U(1)}^2$ &
\\[.5ex]\cline{2-3}&&&\\[-2ex]
$\langle(\rep{16})_{\text{-}3}\rangle \neq 0$
&
$\begin{array}{c}
\frac{1}{8} (\rep{120})_2 +(\rep{16})_{\text{-}3}
\end{array}$
&
$\begin{array}{c}
\frac{1}{8}\left[ (\rep{105})_{1,2}+(\rep{15})_{\text{-}1,6}\right]\\
+\frac{7}{8}(\rep{15})_{2,\text{-}4}
\end{array}$
&
$\begin{array}{lcl}
\vspace{-36pt}
\ass\ass\\
(\rep{1})_{\text{-}3,0}\ass=\as e^T v\\
(\rep{15})_{\text{-}1,\text{-}4}\ass=\as e^{T} (\rep{15})_{2,\text{-}4}
\end{array}$
\\[2ex]
\hline
\end{tabular} 
\caption{
\label{tab:matchC2}
Details of the matching of SO(32)  $\mathbbm{C}^2/\mathbbm Z_2$
models at the spectra level. For each resolution, treated on a separate 
row, we list the orbifold spectrum (second column), the resolution 
spectrum (third column) and the field redefinitions (fourth column) 
necessary to match the two spectra.  As the blow-up induces a gauge 
symmetry breaking the orbifold states are branched accordingly, where 
we do not list those of the adjoint. Afterwards, we make a convenient 
U(1)--basis change. 
Finally, the field redefinition clarifies the matching between
the orbifold and resolution states.
}
\end{center}
\end{table}
}

\def\tbAnomPolyZtwo{ 
\begin{table}[t]\footnotesize
\begin{center}
\begin{tabular}{|c||c|}
\hline&\\[-2ex]
&anomaly polynomials\\[1ex]
\hline\hline&\\[-2ex]
2AI &
\mbox{}\hspace{-12pt}
$\begin{array}{ll}
 \hat I^{het}_8\ass=
\frac{1}{2^8}
\left[(i F_{26})^2 + 2(i F_{2})^2 + 12 (i F)^2 + 6 (iF')^2-R^2 \right]\\[.5ex]
\ass\mbox{}\hspace{20pt}
\left[-2(iF_{26})^2+12(iF_{2})^2+72(iF)^2+84(iF')^2+
192(iF)(iF')-R^2\right]\\[1.5ex]
 \hat I^{uni}_8\ass=
\frac{1}{2^8}
\left[(i F_{26})^2 + 2(i F_{2})^2 + 12 (i F)^2 + 6 (iF')^2-R^2 \right]\\[.5ex]
\ass\mbox{}\hspace{20pt}
\left[-2(iF_{26})^2+12(iF_{2})^2+264(iF)^2+84(iF')^2-
192(iF)(iF')-R^2\right]\\[1.5ex]
 \hat I^{non}_8 \ass= 
\frac{1}{8}(iF)
\Big\{144 (iF)^3+24\left[(iF)+(iF')\right](iF_{2})^2
-144\left[(iF)-(iF')\right] (iF)(iF')
-3(iF) R^2
\Big\}
\end{array}\hspace{-10pt}
$
\\[9ex]
\hline\hline&\\[-2ex]
2BI &
\mbox{}\hspace{-12pt}
$\begin{array}{ll}
 \hat I^{het}_8\ass=
\frac{1}{2^8}
\left[(i F_{20})^2 + 2(i F_{6})^2 + 12 (i F)^2 -R^2 \right]
\left[-2(iF_{20})^2+12(iF_{6})^2+72(iF)^2-R^2\right]\\[1.5ex]
 \hat I^{uni}_8\ass=
\frac{1}{2^8}
\left[(i F_{20})^2 + 2(i F_{6})^2 + 12 (i F)^2 -R^2 \right]
\left[-2(iF_{20})^2+12(iF_{6})^2+72(iF)^2-R^2\right]\\[1ex]
 \hat I^{non}_8 \ass= 
\frac{1}{8}(iF)
\Big\{48 (iF)^3+8 (iF_6)^3+24(iF)(iF_6)^2
-3(iF) R^2
\Big\}
\end{array}\hspace{-10pt}
$
\\[5.5ex]
\hline\hline&\\[-2ex]
2CI &
\mbox{}\hspace{-12pt}
$\begin{array}{ll}
 \hat I^{het}_8 \ass= 
\frac{1}{2^8}\left[
2(i F_{15})^2 + 
12 (i F)^2 + 
80 (iF')^2-R^2 \right]
\left[
72(iF)^2+
480(iF)(iF')+
800(iF')^2-R^2\right]\\[1.5ex]
 \hat I^{uni}_8 \ass= 
\frac{1}{2^8}\left[2(i F_{15})^2 + 12 (i F)^2 + 80 (iF')^2-R^2 \right]
\left[
72(iF)^2-
480(iF)(iF')+
800(iF')^2-R^2\right]\\[1ex]
 \hat I^{non}_8 \ass= 
\frac{1}{8}(iF)
\Big\{48 (iF)^3-1440(iF')^3+4 (iF_{15})^3
+6 \left[(iF)+2(iF')\right](iF_{15})^2
\\[0ex]\ass\mbox{}\hspace{36pt}
-360 \left[(iF)-4(iF')\right](iF)(iF')
-3iF) R^2
\Big\}
\end{array}\hspace{-10pt}
$
\\[8ex]
\hline
\end{tabular} 
\caption{\label{tab:anomalC2}
We give the details of the anomaly cancellation in the blow-up of the
various $\mathbbm{C}^2/\mathbbm Z_2$ SO(32) orbifold models.
For each blow-up we list the orbifold anomaly polynomial $\hat I^{het}_8$
after the gauge symmetry breaking and the resolution
anomaly polynomial, split into the universal $\hat I^{uni}_8$ and non-universal
$\hat I^{non}_8$ parts. Traces are again implicit.}
\end{center}
\end{table}
}

\begin{document}

\thispagestyle{empty}
\vspace*{.2cm}
\noindent
\begin{flushright}
\begin{minipage}{5cm}
HD-THEP-08-9\\
SIAS-CMTP-08-1\\
CPHT-RR002.0108 \\
LPT-ORSAY-08-16
\end{minipage}
\end{flushright}

\vspace*{1.5cm}
\begin{center}
{\Large\bf 
Compact heterotic orbifolds in blow--up} 
\\[.5cm]
{\large Stefan~Groot Nibbelink$^{a}$, 
Denis~Klevers$^b$, 
Felix~Pl\"oger$^{b,c}$, \\
Michele~Trapletti$^{d}$, 
Patrick~K.S.~Vaudrevange$^b$}\\[.5cm]
{\it $^a$ 
Institut f\"ur Theoretische Physik, Universit\"at Heidelberg, 
Philosophenweg 16 und 19, D-69120 Heidelberg, Germany \&
Shanghai Institute for Advanced Study,
University of Science and Technology of China,
99 Xiupu Rd, Pudong, Shanghai 201315, P.R. China
\\[.3cm]
$^b$ 
Physikalisches Institut der Universit\"at Bonn,
Nussallee 12, D-53115 Bonn, Germany
\\[.3cm] 
$^c$ 
Institut f\"ur Chemie und Dynamik der Geosph\"are, 
ICG-1: Stratosph\"are,
Forschungszentrum J\"ulich GmbH, 
D-52425 J\"ulich, Germany
\\[.3cm]
$^d$ 
Laboratoire de Physique Theorique, Bat. 210,
Universit\'e de Paris--Sud, F-91405 Orsay, France \&
Centre de Physique Th\'eorique, \'Ecole Polytechnique,
F-91128 Palaiseau, France }

\vspace{1cm}

{\bf Abstract}\end{center}

\noindent
We compare heterotic string models on orbifolds with supergravity
models on smooth compact spaces, obtained by resolving the
orbifold singularities. Our main focus is on heterotic  $\text{E}_8\times
\text{E}'_8$ models on the resolution of the compact
$T^6/\mathbbm Z_3$ orbifold with Wilson lines. We explain
how different gauge fluxes at various resolved fixed points can be
interpreted in blow down as Wilson lines. Even when such Wilson lines
are trivial from the orbifold perspective, they can still lead to
additional symmetry breaking in blow--up. Full agreement is achieved 
between orbifold and resolved models, at the level of gauge
interactions, massless spectrum and anomaly cancellation. In this
matching the blow--up modes are of crucial importance: they play the 
role of model--dependent axions involved in the cancellation of multiple 
anomalous U(1)'s on the resolution. We illustrate various aspects by 
investigating blow--ups of a $\mathbbm Z_3$ MSSM model with
two Wilson lines: if all its fixed points are resolved simultaneously,
the SM gauge group is necessarily broken. Finally, we explore in 
detail the anomaly cancellation on the complex two dimensional
resolution of $\mathbbm C^2/\mathbbm Z_2$.

\newpage

\section{Introduction}
\label{sc:intro}

Orbifold compactification of the heterotic string \cite{Dixon:1985jw} has
been one of the first approaches to string phenomenology. In the past, 
vast scans of possible 4D models were undertaken with the aim of
reproducing the spectrum and the interactions of the Standard Model 
of particle physics or of a supersymmetric extension of it (MSSM), see e.g.
\cite{Ibanez:1987sn}. The interest in this approach has been recently 
revived with the initial goal to obtain ``orbifold GUTs''~\cite{Altarelli:2001qj} 
from string compactifications~\cite{Kobayashi:2004ya,Forste:2004ie,Hebecker:2004ce,Buchmuller:2005jr,Kim:2006hw}. 
With this technology in mind, the original aim of building a 4D MSSM model 
was re--established, leading to many successful constructions
\cite{Buchmuller:2005jr,Buchmuller:2006ik} that are nowadays some of the best known
string models (for other constructions see e.g.~\cite{Faraggi:1989ka,Blumenhagen:2001te,Dijkstra:2004cc,Braun:2005ux}).

The orbifold constructions have proven to be one of the most
successful  approaches to string phenomenology, yet this approach has
a severe limitation: exact string quantization is only possible on the
orbifold, as it is constructed by combining free conformal field 
theories (CFTs). This means that ``calculability'' is limited to a single
point in the moduli space of the model. This does not mean that away
from the orbifold point one has no control over the resulting 4D model:
we can describe a model away from the orbifold point by giving vevs
to some twisted states. Nevertheless, this extrapolation away from the
orbifold point in moduli space is sensible only if these vevs are
sufficiently small. Otherwise, the standard truncation to the 4D supergravity 
Lagrangian cannot be trusted. However, there are good reasons to
consider big deformations of the orbifold model: having access to
only a limited region in moduli space makes it virtually impossible to
achieve an efficient moduli stabilization mechanism or to study
supersymmetry breaking vacua.

In order to overcome this obstacle, it would be crucial to combine
the model building power of orbifold constructions with an approach 
able to characterize realistic models away from the orbifold point in 
moduli space, i.e. when the orbifold singularities are
resolved~\cite{Aspinwall:1994ev}.
This is the main intention of this paper. To do so, we build on the
results of \cite{Nibbelink:2007rd}, where the resolutions of 
$\mathbbm C^{n}/\mathbbm Z_n$ singularities were considered. 
The freedom in the embedding of the
U(1) bundle on these resolution spaces into the SO(32) gauge group of 
10D heterotic supergravity allowed for the construction of a range of
{\it resolved} models in 4D and 6D.
These models could be matched with the corresponding {\it singular} orbifold models
built by quantizing the heterotic string on $\mathbbm C^{n}/\mathbbm
Z_n$, with $n=2,3$, using the standard CFT techniques.
In this matching it was crucial to ``blow--up'' the orbifold model by
giving a vev to a certain twisted scalar which defines the ``blow--up
mode'' \cite{Hamidi:1986vh}. This matching was refined in \cite{GrootNibbelink:2007ew},
where the issues of multiple anomalous U(1)'s and generalized
Green--Schwarz mechanisms were addressed. Using toric geometry 
\cite{Nibbelink:2007pn} this program can be carried out to a much 
wider class of orbifold singularities.

In the present paper we want to consider the $\text{E}_8\times \text{E}'_8$
heterotic string on the {\it compact} orbifold $T^{6}/\mathbbm Z_3$. 
(For a discussion of the heterotic SO(32) string on such an orbifold, see
e.g.~\cite{Giedt:2003an}.) The compactness of this orbifold is very relevant 
for phenomenology since, apart from a finite 4D Planck mass, it allows us
to include {\it discrete Wilson lines}, which are crucial for model building. 
To prepare for our study of this compact orbifold, we extend the
results of  \cite{Nibbelink:2007rd}  to the $\text{E}_8\times
\text{E}'_8$ heterotic string on $\mathbbm C^3/\mathbbm Z_3$ in 
Section~\ref{sc:C3Z3res}. 
We first recall the resolved geometry and the form of U(1) bundles on
it. After this we consider all possible $\text{E}_8\times
\text{E}'_8$ embeddings of the U(1) flux and describe the resulting five
inequivalent resolved models. We relate each resolution
model to a known heterotic orbifold model by switching on a
certain blow--up mode. We check that the vev of this twisted state is
essentially compatible with F-- and D--flatness. Finally, we explain
how one can use field redefinitions to understand that their spectra
agree in detail.

In Section~\ref{sec:compactorbifold} we construct the resolution of
the {\it compact} $T^6/\mathbbm Z_3$ orbifold by cutting a local patch
around each singularity and replacing it by the resolved space with
U(1) bundles, as described in Section~\ref{sc:C3Z3res}. The matching
in the absence of discrete Wilson lines seems to be a straightforward
extension of the results of Section~\ref{sc:C3Z3res}. However, two
minor complications arise: firstly, the superpotential in the compact
case is generically more complicated than the non--compact one, hence
F--flatness needs to be rechecked. Secondly, it is  possible that
there is a trivial Wilson line between two orbifold fixed points,
which in  blow--up nevertheless leads to a further symmetry breaking.

From the resolution perspective, we interpret discrete Wilson lines 
as the possibility of wrapping different fluxes around each resolved 
singularity. In other words, this freedom can be understood as 
non--trivial transition functions for the gauge backgrounds going 
from one resolved singularity to the other. We study the consistency 
conditions for the transition functions. Furthermore, we show that the 
presence of these transition functions affects the computation of the 
unbroken 4D gauge group and of the localized (twisted) and delocalized
(untwisted) matter spectrum. We conclude this section with two examples: 
the first example illustrates how to embed a discrete Wilson line
on the resolution of $T^6/\mathbbm Z_3$ and exemplifies 
the possibility of having multiple anomalous U(1)'s in compact
blow--ups. The second example demonstrates some of the potential
consequences of blowing up all singularities for semi--realistic
MSSM--like models: contrary to the orbifold theory, in a full resolution
of the model in exam the hypercharge U(1) is necessarily broken.

In Section~\ref {app:C2overZ2} we pass to the study of resolutions of the
$\mathbbm C^2/\mathbbm Z_2$ orbifold and extend the purely topological
approach to the resolution of $T^4/\mathbbm Z_2$ singularities, as
considered in~\cite{Honecker:2006qz}: there it was noted that the 
6D anomaly polynomials of the heterotic orbifold and of the related
smooth models are not the same. We analyze this problem in the same
spirit of Section~\ref{sc:C3Z3res},
matching the models at the level of the gauge interactions and spectra
after giving a vev to a suitable blow--up mode. We show that the
matching of the anomaly cancellations requires carefully considering
the consequences of the field redefinitions that make the U(1) charges of
the models match.

The paper is concluded by Section~\ref{sc:concl}, which summarizes our main findings.

\section{Heterotic $\boldsymbol{\mathbbm C^3/\mathbbm Z_3}$ orbifold
and resolution models} 
\label{sc:C3Z3res}

We consider the heterotic string quantized on the singular
space $\mathbbm M^4\times \mathbbm C^3/\mathbbm Z_3$ 
and on its resolution. We start by giving the geometrical details of the 
$\mathbbm C^3/\mathbbm Z_3$ singularity. 
Then we show how to resolve it and how to construct gauge
fluxes on the resolution. After this study of the geometry, we
consider the heterotic string on the singular space and on the resolution,
leading to 4D heterotic orbifold and resolved models, respectively. Finally, we
investigate how the two classes of models match with particular care
for the anomaly cancellation: we show that, on the orbifold side, the
standard Green--Schwarz mechanism, involving one single universal
axion, is combined with a Higgs mechanism giving rise to the
blow--up. On the resolution, this combination is mapped into a
Green--Schwarz mechanism involving two axions. These are mixtures of
the orbifold axion and of the blow--up mode. This identification is
completed by the observation that the new Fayet--Iliopoulos term
produced on the resolution is nothing else than the (tree--level)
D--term due to the non--vanishing vev of the blow--up mode.

\subsection{Orbifold and blow--up geometry}
\label{sec:geometry}

We start from $\mathbbm C^3$ parameterized by the three complex
coordinates $\tilde{Z}^A$ ($A = 1,2,3$), on which the orbifold
rotation $\Theta$ acts as 
\begin{equation}\label{eq:twist}
\Theta:\,\, \tilde{Z}^A\longmapsto e^{2\pi i \phi_A/3} \tilde{Z}^A,\quad \phi = (1,1,-2)\;.
\end{equation}
The non--compact orbifold $\mathbbm C^3/\mathbbm Z_3$ is obtained by
identifying those points in $\mathbbm C^3$ that are mapped into each
other by $\Theta$. Such a space is singular in the fixed point
$\{0\}$, and is naturally equipped with a K\"ahler potential,
inherited from $\mathbbm{C}^3$,  
\begin{equation}\label{eq:orbKahler}
\mathcal{K}_{\mathbbm{C}^3/\mathbbm Z_3} = \sum_A\bar{\tilde{Z}}^A\tilde{Z}^A\;.
\end{equation}
We can cover $\mathbbm{C}^3/\mathbbm Z_3$-\{0\} by means of
three coordinate patches, defined as
\begin{equation}
U_{(A)}\equiv \lbrace\tilde{Z}\in \mathbbm{C}^3|\tilde{Z}^A \neq 0\,,\, 
0<\mbox{arg}(\tilde{Z}^A)<2\pi/3 \rbrace\;,\;\;\; A =1,2,3 \;.
\end{equation}
It is convenient to choose new coordinates on the orbifold, which allow 
for a systematic construction of a resolution of the singularity as a
line bundle over $\mathbbm{CP}^{2}$.  
In the language of toric geometry \cite{Nibbelink:2007pn,Lust:2006zh},
the $\mathbbm{CP}^2$ is called an exceptional divisor, and it
replaces the singularity in the resolution $\mathcal{M}^3$ of
$\mathbbm C^3/\mathbbm Z_3$. When its volume shrinks to zero, the singularity
is recovered, and the space $\mathcal{M}^3$ 
approaches $\mathbbm C^3/\mathbbm Z_3$ (blow--down). Thus the
blowing--up/down procedure is controlled by the size of the exceptional
divisor.  To make this more explicit we consider the patch $U_{(A)}$,
where $\tilde{Z}^A\neq 0$, and define $z^B\equiv\tilde{Z}^B/\tilde{Z}^A$ 
for  $B\neq A$. To remove the deficit angle of $\tilde{Z}^A$ we
perform the coordinate transformation 
$\tilde{Z}^A\mapsto x \equiv (\tilde{Z}^A)^3$.
In this way the K\"ahler potential becomes 
\begin{equation}
\label{eq:orbKahler2}
\mathcal{K}_{\mathbbm{C}^3/\mathbbm Z_3} = X^{\frac{1}{3}}\;,\,\,\,\,\,
X \equiv \bar{x}(1 + \bar{z}z)^3 x.
\end{equation}  
This change of variables trades the deficit angle for a non--analyticity
in  $\mathcal{K}_{\mathbbm{C}^3/\mathbbm Z_3}$.

\begin{figure}[t]
\hspace{1.4cm}
\begin{center}
\begin{tabular}{c}
\input{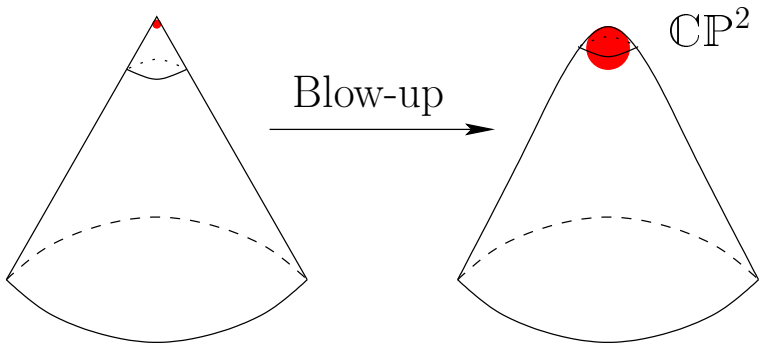}
\end{tabular} 
\end{center}
\caption{\label{fig:2}
The orbifold singularity is cut out locally and a smooth resolution manifold 
(containing an exceptional divisor $\mathbbm{CP}^2$) glued in.}
\end{figure}

A resolution $\mathcal M^3$ of the orbifold is given by considering
the open patches introduced above, equipped with a new K\"ahler
potential \cite{Nibbelink:2007rd} 
\begin{equation}
\mathcal{K}_{\mathcal{M}^3} = \int_{1}^{X}\frac{\mbox{d}X'}{X'} M(X')\;, 
\quad M(X) = \frac{1}{3}(r + X)^{\frac{1}{3}}\;,
\end{equation}
that is Ricci--flat and matches the orbifold K\"ahler potential in the
$r\rightarrow 0$ limit. In this limit the curvature vanishes for
points where $x\neq 0$, whereas for $x=0$ it diverges. Moreover, it
vanishes for any value of $r$ when $|x|\rightarrow\infty$. Therefore,
blowing up means that the orbifold singularity is replaced by 
the smooth compact $\mathbbm{CP}^{2}$ that shrinks to zero as
$r\rightarrow 0$ (the situation is illustrated in Figure~\ref{fig:2}).

\subsection{Gauge fluxes wrapped on the orbifold and the resolution}
\label{sc:gaugefluxes}

When defining the heterotic string on $\mathbbm C^3/\mathbbm Z_3$, 
the 10D gauge group $\text{E}_8 \times \text{E}'_8$ \footnote{We restrict
to this case as the $\text{SO(32)}$ theory was considered in
\cite{Nibbelink:2007rd,GrootNibbelink:2007ew}.}  is broken  
by the orbifolding procedure.
We can understand this breaking from an effective field theory perspective:
let $i \mathfrak{A}$ be the one--form gauge field,  with values in
$\text{E}_8 \times \text{E}'_8$, and let  $i \mathfrak{F}$ be its field
strength. Moreover, define $H_I$, for $I=1,\ldots,16$, as the basis
elements of the Cartan subalgebra of $\text{E}_8 \times \text{E}'_8$.
In a given coordinate patch with local coordinates $z, x=|x|e^{i\phi}$,
the orbifold action $\Theta$ is realized as $\phi\rightarrow\phi+2\pi$. 
On the orbifold there can be non--trivial orbifold boundary conditions
for $\mathfrak{A}$ 
\begin{equation}
\label{orbbreak}
i \mathfrak{A}(\Theta\, \tilde{Z}) =
i \mathfrak{A}(z,|x|,\phi+2\pi) = U i \mathfrak{A}(z,|x|,\phi)U^{-1}\;, 
\end{equation} 
where $U=e^{2\pi i (V^I_{orb} H_I)/ 3}$ and $V_{orb}$ is a vector in the
$\text{E}_8 \times \text{E}'_8$ root lattice. These boundary
conditions induce a gauge symmetry breaking, precisely to those 
$\text{E}_8 \times \text{E}'_8$ algebra elements with root vectors $p$
such that $p \cdot V_{orb} = 0\text{ mod } 3$. Our normalization of the
orbifold gauge shift vector $V_{orb}$ differs by a factor $1/3$ from
the common one; our convention avoids such an additional
factor when we make identification with gauge bundle fluxes below.

The non--trivial orbifold boundary conditions can be reformulated in
terms of fields with trivial ones, but having a non--zero constant gauge
background. The existence of this non--vanishing gauge flux, localized
at the singularity, should become ``visible'' on the resolution. To
obtain a matching of orbifold models with models built on the resolved
space, we consider the possibility of a gauge bundle wrapped around
the resolution. In general such a bundle has structure group
$\mathcal{J}$ embedded into $\text{E}_8 \times \text{E}'_8$.  This
embedding breaks the 10D gauge group $\text{E}_8 \times \text{E}'_8$ 
to the maximal subgroup $H \subset \text{E}_8 \times
\text{E}'_8$ that commutes with $\mathcal{J}$.
We therefore expand the 10D gauge field strength 
$i\mathfrak{F}=i\mathcal{F}+iF$ around the internal background
$i\mathcal{F}$, living in the algebra of $\mathcal{J}$, in terms of the
4D gauge field strength $iF$, taking values in the algebra of $H$. 
To preserve $N=1$ supersymmetry in four dimensions, the bundle field
strength has to satisfy the Hermitian Yang--Mills
equations~\cite{Candelas:1985en}\footnote{Here we ignore loop
corrections to these equations, discussed in \cite{Blumenhagen:2005ga}. We
will return to this point later in the paper. }   
\begin{equation}\label{eq:YM}
\mathcal{F}_{AB}=0\;,\quad \mathcal{F}_{\bar{A}\bar{B}}=0\;,
\quad G^{A\bar{A}}\mathcal{F}_{A\bar{A}}=0\;,
\end{equation}
where $G^{A\bar{A}}$ denotes the inverse Hermitian metric of
$\mathcal{M}^3$. One further (topological) consistency requirement
follows from the integrated Bianchi--identity of the two--form $B$ of
the supergravity multiplet: 
\begin{equation}\label{eq:Bianchi1}
\int_{\mathcal{C}^4}\left(\text{tr}\mathcal{R}^2 - \text{tr}(i\mathcal{F})^2 \right) = 0\;,
\end{equation}
for all compact four--cycles $\mathcal{C}^4$ of the resolution and $\mathcal{R}$ 
denotes the curvature of the internal space $\mathcal{M}^3$. This
condition is crucial to ensure that the effective four dimensional
theory is free of non--Abelian anomalies~\cite{Witten:1984dg}. The
resolution space $\mathcal{M}^3$ only contains a single compact four--cycle, the
$\mathbbm{CP}^2$ at the resolved singularity, leading to a single consistency condition.

We give two examples of bundles on the resolution that
satisfy~(\ref{eq:YM}) and (\ref{eq:Bianchi1}). The simplest
construction of such a bundle is the \textit{standard embedding} 
(to which we refer as ``AS'') with
the gauge connection taken to be equal to the spin connection
\cite{Candelas:1985en}. In terms of the curvature this means 
$i \mathcal{F}=\mathcal{R}$. Since $\mathcal{R}\in$ SU(3), this
describes an SU(3) bundle, embedded into $\text{E}_8
\times\text{E}'_8$, leading to the 4D gauge group
$H=\text{E}_6\times\text{E}'_8$. However, in this paper we mainly
focus on the U(1) gauge bundle with field strength
\begin{equation}\label{eq:Uonebundles1}
i \mathcal{F} = 
 \Big(\frac r{r + X} \Big)^{1-\frac 1n}
\Big( 
\bar e e - \frac {n-1}{n^2} \, \frac 1{r+X}\, \bar \epsilon \epsilon
\Big)\;, 
\end{equation}
see \cite{Nibbelink:2007rd} for notational conventions. Such a 
bundle can be embedded into $\text{E}_8\times\text{E}'_8$ as 
\begin{equation}\label{eq:Uonebundles2}
i \mathcal{F}_V = i \mathcal{F}\, H_V\;,
\end{equation}
where we use the notation $H_V \equiv V^I H_I$. Since the bundle is
only well--defined if its first Chern class, integrated over all
compact two--cycles, is integral, an extra consistency requirement
arises for  the vector $V^I$. For the two--cycle $\mathbbm{CP}^1$ at
$x=0$, 
\begin{equation}
\frac{1}{2\pi i} \int_{\mathbbm{CP}^1} i \mathcal{F}_V = V^I H_I
\end{equation}
must be integral for all $\text{E}_8 \times\text{E}'_8$ roots. 
This implies that $V$ has to be an $\text{E}_8 \times\text{E}'_8$
root lattice vector itself. The two--form $\mathcal{F}$ is regular
everywhere for $r\neq 0$. In the blow--down limit $r\rightarrow 0$, it
is zero for $x\neq 0$ and it diverges for $x=0$, in such a way that
the integral above remains constant. This means that the bundle is
``visible'' as a two--form only in the blow--up, but in the blow--down its
physical effect is not lost because the gauge flux gets localized in
the fixed point. In this sense, this bundle is exactly the counterpart
of the orbifold boundary conditions discussed above.

\tbVorbClass

\subsection{Classifying orbifold and smooth line bundle models}
\label{sec:classification}

The heterotic string on the $\mathbbm C^3/\mathbbm Z_3$ is 
specified by the orbifold gauge shift vector $V_{orb}$ defined
in~(\ref{orbbreak}). The freedom in the choice of $V_{orb}$ is
constrained by modular invariance of the string partition function:
\begin{equation}
V_{orb}^2 = 0 ~\text{mod}~6\;.
\end{equation}
There are only five inequivalent shift vectors, each of them giving
rise to a different orbifold model. In Table~\ref{orbshift} we list the possible
$V_{orb}$ together with the gauge groups surviving the orbifold projection.
Using the standard CFT procedure it is possible to compute the spectra of these
models. They are listed in the second column of Table~\ref{tab:match}. The spectra
are given with the multiplicity numbers with which the various
states contribute to the 4D anomaly polynomial localized in the singularity.
Thus, these numbers can be fractional if the corresponding states
are not localized in the $\mathbbm C^3/\mathbbm Z_3$ singularity. The
untwisted states have multiplicities that are multiples of
$3/27$, because the compact orbifold $T^6/\mathbbm Z_3$ has 27 singularities and 
untwisted states come with multiplicity three. On the other hand,
these multiplicities are integers for localized (i.e. twisted) states.

The blow--up model is completely specified by the way how the
gauge flux is embedded in $\text{E}_8 \times\text{E}'_8$, i.e. by the 
vector $V$. The Bianchi identity integrated over $\mathbbm C \mathbbm P^2$ yields 
the consistency condition
\begin{equation}\label{eq:Bianchi2}
V^2 = 12
\end{equation}  
and enormously constrains the number of possible models. All solutions
to~(\ref{eq:Bianchi2}) together with the corresponding gauge groups
are given in Table~\ref{tab:E8shifts}. The chiral matter content is
determined by the Dirac index theorem that for U(1) bundles takes the
form of a multiplicity operator 
\begin{equation}\label{eq:index2}
N_V = \frac{1}{18} ( H_V)^3 - \frac{1}{6} H_V\;, 
\end{equation}
see \cite{Nibbelink:2007rd} for details. The computation of the
spectra for each of the U(1) embeddings shows that there are in fact
only five inequivalent models amongst them. We distinguish them by
their chiral spectra, which are given in the third column of
Table~\ref{tab:match}. 

\tbVblowClass

\subsection{Matching orbifold and blow--up models}
\label{sec:localmatching}

We now want to investigate the matching between the heterotic orbifold
and the blow--up models discussed in the previous section. This
matching can be considered at various levels and we begin with some
simple observations before entering more subtle issues. 

The first basic observation was made in (\ref{orbbreak}):
the embedding of the orbifold rotation in the gauge bundle
(via the shift $V_{orb}$) can be seen as the presence of
a gauge flux localized in the singularity. On the resolution, such
a gauge flux appears, and it is immediate to identify
\begin{equation}
\label{matchV}
\frac{1}{3}V^I H_I = \frac{1}{2\pi i}\int_\mathbbm{C} i\cF_V
\rightarrow  \frac{1}{3}V^I_{orb} H_I.
\end{equation}
The integration above is made on the variable $x$ defined in~(\ref{orbbreak})
in such a way that the integral can be immediately read as a contour
integral of $\cA$ around the phase $\phi$ of $x$ or, in other words,
precisely as the Wilson line associated with $V^I_{orb} H_I$
in~(\ref{orbbreak}).

This basic observation is corroborated by the fact that any blow--up shift 
$V$, listed in Table \ref{tab:E8shifts}, is modular invariant,  because 
$V^2=12=0$ mod $6$. At first sight the reverse, any 
orbifold shift $V_{orb}$, classified in Table \ref{orbshift}
corresponds to a blow--up, does not seem to be true. However, we should
take into account that  two orbifold shift vectors are equivalent,
i.e.\ lead to the same model, if: i) they differ by $3\Lambda$ where
$\Lambda$ is any element of the root lattice of 
$\text{E}_8\times\text{E}'_8$, ii) they differ by sign flips of an even
number of  entries, or iii) are related by Weyl reflections. 
By suitable combining these operations one can show that all blow--up
vectors of Table \ref{tab:E8shifts} can be obtained from the orbifold
shifts in Table \ref{orbshift}. (Only the first model in
Table~\ref{orbshift}, characterized by  the zero vector $(0^8)(0^8)$,
does not have a blow--up counterpart in Table \ref{tab:E8shifts}.)  
This leads to a direct matching between orbifold and blow--up models. 
Using the notation from the Table~\ref{orbshift} and
\ref{tab:E8shifts}, we match model B with BI, model D with DI. We also
see that even though CI and CII are different blow--up models, they
correspond to the same orbifold theory C. The same applies to the U(1)
bundle model AI and the standard embedding model AS (introduced in
Section~\ref{sc:gaugefluxes}): They are both related to orbifold model A.

Given the matching at the level of the gauge bundles, we can pass to
checks at the level of the 4D gauge groups. A quick glance over the
Table~\ref{orbshift} and~\ref{tab:E8shifts} shows that their gauge
groups are never the same. This is easily explained from the orbifold
perspective: the blow--up is generated by a non--vanishing 
vev of some twisted state, the so--called \textit{blow--up mode}
\cite{Font:1988tp}.  As all twisted
states are charged, this vev induces a Higgs mechanism accompanied with
gauge symmetry breaking and mass terms. It is not difficult to see
from these tables that all non--Abelian blow--up gauge groups can be
obtained from the orbifold gauge groups by switching on suitable vevs
of twisted states.

Even after taking symmetry breaking, i.e. the branching of the
representations of the orbifold state, into account the spectra of the
orbifold models still do not agree with the ones of the resolved models: singlets w.r.t.\ 
non--Abelian blow--up groups, and some vector--like states are
missing. Moreover, the U(1) charges of localized states do not
coincide with the ones expected from the branchings. This can be
confirmed from Table~\ref{tab:match}: for each model we give the
orbifold spectrum (second column) and the resolution spectrum (third
column).

All these differences can be understood by more carefully taking into
account the possible consequences of a twisted state's vev $v$. 
After branching, this field is a singlet of the non--Abelian gauge
group. In the quantum theory this means that the corresponding chiral
superfield $\Psi_q$ with charge $q$ under the broken U(1) never
vanishes. Hence, it can be 
redefined as $\Psi_q = v e^T$, where $T$ is a new chiral
superfield taking unconstraint values. As it transforms as  an axion 
\begin{equation}\label{eq:axiontransf}
T \longrightarrow T + i q \phi\;,
\end{equation}
under a U(1) transformation with parameter $\phi$, it is neutral.
Hence, it is not part of the charged chiral spectrum computed using the
Dirac index~\eqref{eq:index2} on the resolution. In addition, we can use this axion
chiral superfield $T$ to redefine the charges of other twisted states 
(see the last column of Table~\ref{tab:match}) so that all U(1)
charges of the twisted states agree with the ones of the localized resolution
fields. For models CII and DI one needs in addition to change the U(1)
basis when identifying the orbifold and blow--up states, if one enforces
that the field getting a vev is only charged under the first blow--up
U(1) factor.

\tbMatchingZthree

\noindent
Finally, the remaining states that are missing in blow--up
(referred to in Table~\ref{tab:match} with a superscript $m$) 
have Yukawa couplings with the blow--up mode, so that they get a
mass term in the blow--up.
Taking all these blow--up effects into account shows that the
spectra of the blown--up orbifold and resolution models become perfectly identical.

\subsection{F-- and D-- flatness of the blow--up mode}
\label{sec:Fflatness}

In the matching of heterotic orbifold models with their resolved 
counterparts we assumed that a single twisted field of the orbifold model was
responsible for generating the blow--up. No other twisted or untwisted
states attained non--vanishing vevs. However, in order to obtain a
supersymmetric configuration, we have to pay attention  to possible
non--vanishing F--terms arising from the non--zero vev. The analysis
of the F--flatness for a superpotential $\mathcal W$ is rather involved in
the context of heterotic orbifold model building, because in principle
it contains an infinite set of terms with coefficients determined by
complicated string amplitudes. In practice string selection rules can
be used to argue that a large class of these coefficients vanishes
identically, while the others are taken to be arbitrary
\cite{Kobayashi:1990mc,Lebedev:2007hv,Choi:2007nb}.

Our assumption above that only a single twisted superfield
has a non--vanishing vev greatly simplifies the F--flatness analysis: 
non--vanishing F--terms can only arise from terms in the superpotential 
that are at most linear in fields having zero vevs. As in most of the
cases the vanishing vev fields form non--Abelian representations
gauge invariance of the superpotential implies that they cannot appear
linearly. This means that the complicated analysis of the
superpotential involving many superfields often reduces to the analysis of a
complex function of a single variable. In what follows we show that all
the blow--ups described previously are F--flat and therefore
constitute viable resolutions of orbifold models.

Non--vanishing D--terms can only arise under the following conditions 
\cite{Buccella:1982nx}: let $\varphi_q$ be the scalar component of the only 
superfield $\Psi_q$ that acquires a vev $\langle\varphi_q\rangle$ as discussed above. 
The D--terms are proportional to $D^a \sim \bar{\varphi}_qT^a \varphi_q$, 
where $T_a$ are the generators of the orbifold gauge group $G$. Therefore, 
certainly all D--terms corresponding to the generators $T_a$ that annihilate
$\langle\varphi_q\rangle$ vanish. They generate the {\em little group} $H$ of 
gauge symmetries unbroken by the vev. Consequently, non--vanishing
D--terms are only possible for the generators $T^a$ of the coset $G/H$. Under
an infinitesimal gauge transformation with parameter $\epsilon$ the
D--terms transform as 
$D^a \rightarrow D^a + \bar{\varphi}_q[\epsilon, T^a] \varphi_q$. This means
that for all generators $T^a$ which do not commute with all generators
of $G/H$, we can find a gauge such that the $D^a$'s associated to them
vanish. But since $(D^a)^2$ defines a gauge invariant object, all these
$D^a$ have to vanish in any gauge. The only possibly non--vanishing
D--terms correspond to the Abelian subgroup of the coset $G/H$. As we
explain in the next subsection, precisely those D--terms, which are
associated with anomalous U(1)'s on the resolution $\mathcal{M}^3$, are 
non--vanishing. Apart from this subtle issue, D--flatness is automatically 
guaranteed.

\subsubsection*{Matching A$\rightarrow$AS by a vev of
  $\boldsymbol{3}_H(\boldsymbol{1},\boldsymbol{\overline{3}};\boldsymbol{1})$}

We begin our analysis with the standard embedding defined by a
gauge bundle with $i\cF = \cR$. Because both transform by
conjugation, $i\cF \rightarrow g^{-1} i\cF g$ and 
$\cR \rightarrow g_H^{-1} \cR g_H$ under gauge and internal local
Lorentz (holonomy) transformations, respectively, we know that these
gauge transformations are identified: $g = g_H$. This fact will help
us to identify the blow--up field in the following.

To reconstruct the corresponding blow--up mode we need a field that 
transforms under both of these transformations. In orbifold model A 
the only candidates for this are the three triplets
$3 (\boldsymbol{1},\boldsymbol{\overline{3}};\boldsymbol{1})$. The
multiplicity three is due to internal oscillator excitations of
these states, i.e.\ these states form a triplet under the R--symmetry
group SU(3)$_R$ as it commutes with the $\mathbbm Z_3$ orbifold
holonomy, which in turn is proportional to the identity. A more 
precise way of referring to these states is therefore: 
$\boldsymbol{3}_R(\boldsymbol{1},\boldsymbol{\overline{3}};\boldsymbol{1})$;
we can view them collectively as a 3$\times$3 matrix
$G^\alpha{}_i$, where $\alpha$ is the SU(3)$_R$ index and $i$ the
SU(3) index\footnote{We use the subscript notation to indicate that it
is in the complex conjugate representation.}. The SU(3) gauge and
SU(3)$_R$ R--symmetry groups act on it as $G \rightarrow g_R G g^{-1}$. 
Since any complex matrix $M$ can be written as a product $M=UH$ of
a unitary matrix $U$ and a Hermitian matrix $H$, which in turn can be
diagonalized by another unitary matrix $V$ as $H=VDV^{-1}$ with $D$ a
real diagonal matrix, we can use the gauge and R--symmetry 
transformations to bring $G$ in a real diagonal form. If we choose the vev 
matrix $G$ to be proportional to the identity: $G = v \Identity$, we
find that only a diagonal gauge and R--symmetry transformation
preserves this vev. This means that in the blow--up the symmetry 
$\text{SU}(3)_R \times \text{SU}(3)$ is broken to the diagonal 
$\text{SU}(3)$ subgroup (with $g = g_R$). Comparing to the standard 
embedding on the resolution (with $g = g_H$), the vev of $G$ changed 
the $\mathbbm Z_3$ orbifold holonomy to the full SU(3) holonomy 
of a Calabi--Yau.

To understand whether such a vev for $G$ is possible, we need to
analyze the superpotential of the theory. Because
the $\mathbbm Z_3$ action is proportional to the identity, the
$\mathbbm C^3/\mathbbm Z_3$ orbifold theory is left invariant by any
unitary transformation U(3)$_R$ of the internal coordinates, there is no
superpotential involving only $G$: the SU(3)$\times$SU(3)$_R$
invariance requires such superpotential to be a function of $\det G$,
but that is not invariant under U(1)$_R$. (Similar arguments for the
compact $T^6/\mathbbm Z_3$ allow a cubic superpotential for $G$ in
that case~\cite{Hamidi:1986vh}.) Hence, any vev for $G$ defines a
F--flat direction. However, it can be seen that D--flatness requires 
it to be of the form above~\cite{Hamidi:1986vh}: The SU(3) D--terms
correspond to the traceless part of the matrix
$D^i{}_j = \bar{G}^i{}_\alpha G^\alpha{}_j$. In the diagonal form the
matrix $G$ has the vevs $v_1$, $v_2$ and $v_3$ as its diagonal
elements. This means that $D$ is a diagonal matrix with entries
$v_1^2$, $v_2^2$ and $v_3^2$. But this has a non--vanishing traceless
part unless all vevs are equal.

\subsubsection*{Matching {A$\rightarrow$AI} by a vev of
$(\boldsymbol{27},\boldsymbol{1};\boldsymbol{1})$}

In this case the blow--up mode is in the twisted state
$(\boldsymbol{27},\boldsymbol{1};\boldsymbol{1})$, and the relevant
part of the superpotential are formed from its cubic E$_6$ invariant 
\begin{equation}
\mathcal{W} \sim 
[(\boldsymbol{27},\boldsymbol{1};\boldsymbol{1})]^3 + \ldots. 
\end{equation}
Here, the notation $\sim$ indicates that we only give the lowest order 
gauge invariant structure ignoring its (order one) coefficient, and 
$+\ldots$ means that there is a whole power series of 
this invariant, restricted by some string selection rules. Such
a superpotential is always F--flat. To see this, consider  the
branching of the twisted state  
\( 
\boldsymbol{27}\rightarrow 
\boldsymbol{16}_{-1} + \boldsymbol{10}_2 + \boldsymbol{1}_{-4}
\) 
due to its own vev $v$ that breaks 
$\text{E}_6\rightarrow \text{SO(10)} \times \text{U(1)}$. 
In terms of $\text{SO(10)} \times \text{U(1)}$--representations the invariant reads as
\begin{equation}\label{E6invariant}
[\boldsymbol{27}]^3 = 
\boldsymbol{16}_{-1}\times\boldsymbol{16}_{-1}\times\boldsymbol{10}_2 +
\boldsymbol{10}_2\times \boldsymbol{10}_2\times \boldsymbol{1}_{-4}\;.
\end{equation}
Since the $\boldsymbol{1}_{-4}$ represents the blow--up mode, and hence
by definition the $\boldsymbol{16}_{-1}$ and $\boldsymbol{10}_2$ have
zero vev, F--flatness is automatically guaranteed.  It is clear from this
decomposition of the cubic invariant that the $\boldsymbol{10}_2$
becomes massive and decouples, while the  $\boldsymbol{16}_{-1}$ stays
strictly massless. This is in agreement with the blow--up spectrum given in
Table~\ref{tab:match}.

\subsubsection*{Matching B$\rightarrow$BI by a vev of 
$(\rep{1},\rep{3};\rep{1},\rep{3})$}

In orbifold model B the only twisted state is a
$(\rep{1},\rep{3};\rep{1},\rep{3})$--plet, hence it is the
only possible blow--up mode. Like the blow--up mode in the case of standard
embedding AS discussed above, this blow--up mode defines a 3$\times$3
matrix denoted by $C$. Gauge transformations with $g\in$ SU(3) and $g'\in$ SU(3)$'$ act via 
left and right multiplication $C \rightarrow g C g'$. The relevant part 
of the superpotential is therefore also very similar 
\begin{equation}
\mathcal{W} \sim \det C + \ldots. 
\end{equation}
As the two SU(3)'s are independent, we can again assume that the matrix
$C$ is diagonalized. To obtain the appropriate symmetry breaking  
$\text{SU}(3) \rightarrow U(2)$ of both SU(3)s, only one of the three
diagonal elements has a non--vanishing vev.  This is a very different
orientation of the vev as compared to the standard embedding. 
Expanding the superpotential around this vev, shows that
the state  $(\rep{1},\rep{2};\rep{1},\rep{2})^{m}_{0,0}$ becomes massive.

\subsubsection*{Matching C$\rightarrow$CI by a vev of
$(\rep{1},\rep{1})_{-4,0}$}

In orbifold model C we can construct gauge invariant structures for the 
superpotential only by combining the states $(\rep{1},\rep{1})_{-4,0}$, 
and $(\rep{1},\rep{14})_{2,0}$
\begin{equation} \label{modelCcoupling}
\mathcal{W} \sim (\rep{1},\rep{1})_{-4,0} [(\rep{1},\rep{14})_{2,0}]^2
+ \ldots. 
\end{equation}
Since the CI blow--up is realized by giving a vev to 
the orbifold state $(\rep{1},\rep{1})_{-4,0}$,
which is always coupled to pairs of $(\rep{1},\rep{14})_{2,0}$'s in
the superpotential, this vev defines a flat direction of the potential
and a mass term for the $(\rep{1},\rep{14})_{2,0}$ is generated, 
provided that we perform the field redefinition indicated in 
Table~\ref{tab:match}. Hence, this state decouples.

\subsubsection*{Matching C$\rightarrow$CII by a vev of $(\rep{1},\rep{14})_{2,0}$}

The CII blow--up is obtained when $(\rep{1},\rep{14})_{2,0}$ gets a
vev. Naively one expects that a vev for this state would lead to a
symmetry breaking $\text{SO}(14) \rightarrow \text{SO}(13)$, but this
is not in agreement with Table~\ref{tab:match}. To understand what is
happening we have to consider the possible orientations of
such a vev $C_m$, where $m$ denotes the SO(14) vector index. Since all
states are chiral multiplets, we cannot use the real group SO(14) to
put the vev in a single component. Indeed, writing 
$C_m = R_m + i\, J_m$ where $R_m$ and $J_m$ are real, we see
that one can use a SO(14) transformation to obtain $R=(r, 0^{13})$. This
orientation is left invariant by SO(13) subgroup. This subgroup can be
used to bring $J$ to the form $J=(j_1,j_2,0^{12})$. Hence, for generic
values of $r, j_1$ and $j_2$ only the SO(12) subgroup is left
unbroken, as Table~\ref{tab:match} implies. Furthermore, the superpotential 
contains again the coupling~(\ref{modelCcoupling}).
To have the auxiliary
component of the superfield  $(\rep{1},\rep{1})_{-4,0}$ vanishing in
extremum, the vev of the SO(14) invariant 
\begin{equation}
C^T C = r^2 - j_1^2 - j_2^2 + 2 i\, r j_1
\end{equation}
has to vanish. The only non--vanishing solution has: $j_1 = 0$ and 
$j_2^2 = r^2 =v^2$. This vev 
induces a mass by pairing up one of the singlets from the branching of
$\rep{14} \rightarrow \rep{12} + \rep{1} + \rep{1}$ with the singlet
already present in the orbifold spectrum, see Table~\ref{tab:match}.

Notice that there is a third field in model $C$ that could have a 
non--zero vev, the R--symmetry triplet $\rep{3}_R(\rep{1};\rep{1})_{0,4}$. 
This superfield cannot appear in any superpotential by itself, this 
means that any vev for this superfield leads to a supersymmetric 
configuration. Nevertheless, we do not have any candidate for a 
$\text{U}(1)$ gauge configuration on the resolution that corresponds to this vev.

\subsubsection*{Matching D$\rightarrow$DI by a vev of $(\crep{9},1)_{-4/3}$}

Finally, we consider the orbifold model D. As it has only one charged 
twisted state $(\crep{9},1)_{-4/3}$ it is not possible, due to the
string selection rules, to write down any superpotential with terms at
most linear in the other fields. Thus, it can attain any  vev leading
to the symmetry breaking as described in Table~\ref{tab:match}.

\subsubsection*{Other gauge bundles on the resolution?}

The list of possible vevs of twisted states of a given heterotic
orbifold model is exhausted only for the last case, model D. The
other models allow other blow--ups in principle:

First of all, model $C$ also has an SU(3)$_R$ triplet of scalars, there
is no obvious reason why one of them cannot have a non--vanishing
vev. Model B allows for other possible orientations for the vev of the
$(\rep{1},\rep{3};\rep{1},\rep{3})$ state, because it defines a
$3\times 3$--matrix with three eigenvalues. Thus, in general one should 
allow for blow--ups defined by a multitude of vevs for
possibly all the twisted states that a given orbifold model possesses.

Since our classification of Abelian gauge bundle models on
the resolution of $\mathbbm C^3/\mathbbm Z_3$ is complete, and we have
identified the blow--up modes in the various heterotic theories leading 
to these models, we conclude that other (multiple field) vevs
correspond to non--Abelian bundles on the resolution. Aside from the
standard embedding, model AS, their classification is beyond the scope
of this paper.

\subsection{Multiple anomalous U(1)'s on the blow--up}
\label{sec:anomalies}

In this section we investigate the anomaly cancellation and D--flatness 
on the resolution $\mathcal{M}^3$. We find that there can be at most 
two anomalous U(1)'s, and that their cancellation involves two axions 
\cite{Peccei:1977hh,Peccei:1977ur}, the \textit{model--independent} and 
a \textit{model--dependent} one.\footnote{For
a recent review on axions from string theory see \cite{Svrcek:2006yi}.}
We show that the counterpart of such an anomaly cancellation, from the 
orbifold perspective, is a mixture of the standard orbifold Green--Schwarz
mechanism and the Higgs mechanism related to the blow--up mode.
From this we deduce relations between the two axions and their 
orbifold counterparts, namely, the universal axion of heterotic orbifold 
models and a second field related to the blow--up mode. Finally we discuss 
the issue of D--flatness of the resolution. We show that the blow--up 
is not along a D--flat direction. Rather, in the blow--up a constant 
D--term is produced, which is matched, from the resolution perspective, 
with the appearance of a new Fayet--Iliopoulos term due to the presence 
of two, rather then one anomalous U(1)'s.

\subsubsection*{Anomalous U(1)'s on the resolution: the axions}

We deduce the 4D anomaly polynomial $\hat I_6$ from dimensional reduction of 
the 10D one, $\hat I_{12}$. For notational convenience we absorb some
factors $2 \pi i$ in the definition of the anomaly polynomial: 
$\hat I_{2n+2} = (2\pi i)^n I_{2n +2}$. The anomaly polynomial factorizes as
$\hat I_{12} = X_4 \cdot X_8$, where~\cite{Green:1984sg,Witten:1987}
\begin{equation}
X_4 = \mbox{tr}\, \mathfrak{R}^2 - \mbox{tr} (i \mathfrak{F})^2\;,
\end{equation}  
\begin{equation}
X_8 = \frac{1}{96}\left[
\frac{\Tra (i\frF)^4}{24} - 
\frac{(\Tra (i\frF)^2)^2}{7200}  - 
\frac{\Tra (i\frF)^2 \tr \frR^2}{240}  +
 \frac{\tr \frR^4}{8}  + 
 \frac{(\tr \frR^2)^2}{32}  \right]\;, 
\end{equation}
with $\mathfrak{R}$ denoting the 10D curvature.
The trace $\mbox{tr}$ in the ``fundamental'' of $\text{E}_8 \times\text{E}'_8$ 
is formally defined via $\mbox{tr} =\frac{1}{30}\mbox{Tr}$, $\mbox{Tr}$ 
being the standard trace in the adjoint representation. From $\hat I_{12}$ the 
4D anomaly polynomial $\hat I_6$ can be derived via an integration over the 
resolution manifold. The integration will be performed after inserting 
the expansions $\frR = R+\cR$ and $i\frF = i F+i\mathcal{F}_V$ and 
splitting the forms $X_4$ and $X_8$ according to
\begin{equation}
\hat I_{12} = X_{4,0}\, X_{2,6} + X_{2,2}\, X_{4,4} + X_{0,4}\, X_{6,2} \;,
\end{equation}
where we read $X_{a,b}$ as an $(a+b)$-form with $a$ indices in the 6D 
internal and $b$ indices in the 4D Minkowski space. Since the backgrounds 
are such that the $H_3$ Bianchi Identity is fulfilled, the 4D anomaly 
polynomial $\hat I_6$ is written as the factorized sum
\begin{equation}
\hat I_6 \equiv \frac1{(2\pi i)^3} \int_{\mathcal{M}^3} \hat I_{12} = 
\frac{1}{(2 \pi i)^3}\int_{\mathcal{M}^3}\left( X_{2,2}\, X_{4,4} + X_{0,4}\, X_{6,2}\right) \;.
\end{equation} 
Inserting the expressions for $X_{2,2}$ and $X_{0,4}$ in terms of the 
field strengths and rearranging the terms on the right hand side yields
\begin{equation}\label{eq:factanom}
\hat I_{6} = \hat I^{uni}_6 + \hat I^{non}_6\quad \text{with}\quad
\hat I^{uni}_6 = X^{uni}_2\cdot X_{0,4}\quad \text{and}\quad  
\hat I^{non}_6  = X^{non}_2 \cdot X^{non}_4 \;,  
\end{equation}
where $X^{non}_2 = -2\,\text{tr}[H_V i F],$
$X_{0,4} = \text{tr}R^2 - \text{tr}(i F)^2,$ and 
\begin{equation}
X^{uni}_2 = \frac{1}{(2\pi i)^3}\int\limits_{\mathcal{M}^3}X_{6,2}
\;,\quad
 X^{non}_4 = \frac{1}{(2\pi i)^3}\int\limits_{\mathcal{M}^3}i \mathcal{F} X_{4,4}\;. 
\end{equation}
Now the integration is performed and results in
\begin{equation}
X^{uni}_2 = - \frac{1}{96}\text{Tr} \left[\left(\frac{1}{18} H_V^3 - \frac{1}{5}
H_V\right)(iF) \right] \;,
\end{equation}
\begin{equation}
X^{non}_4 = -\frac{1}{192}\left[ \text{Tr} \left[\left(\frac{1}{6} H_V^2 - 
\frac{1}{5}\right)(i F)^2 \right] - 
\frac{1}{3\cdot 30^2}\left(\text{Tr}[H_V (i F)]  \right)^2 - \tr R^2\right]\;. 
\end{equation}
These equations describe how the 4D anomaly $\hat I_6$ can be written as 
a sum of two factorized parts, a \textit{universal} part $\hat I^{uni}_6$ 
and a \textit{non--universal} part $\hat I^{non}_6$. Since both parts are 
proportional to $\text{tr}(i F)$, they are non--vanishing only for 
anomalous U(1)--factors. As we started from an anomaly free theory in 10D
the 4D Green--Schwarz mechanism will cancel the two summands in 
$\hat I_6$ by 
two axions. The \textit{universal} anomaly is canceled by the anomalous 
variation of the \textit{model--independent} axion and the 
\textit{non--universal} part by the \textit{model--dependent} axion, as 
shown in the following. Therefore on the resolution of 
$\mathbbm{C}^3/\mathbbm Z_3$ there can be at most two anomalous U(1)'s.
In Table~\ref{tab:anomal}, we give the anomaly terms for each of the
$\mathbbm C^3/\mathbbm Z_3$ models.

The 4D anomaly must be canceled by the anomalous variation of the 10D
two-form $B_2$, which can be expanded as
\begin{equation}
B_2 = b_2 + i \cF b_0 + \omega_2 B_0\;.
\end{equation}
The K\"ahler form $\omega_2$, obtained from the K\"ahler potential,  
and the U(1) gauge bundle field strength $i\mathcal{F}$ are harmonic 
two--forms on the resolution $\mathcal{M}^3$. Thus $b_0$ and $B_0$ are 
4D massless scalars. In addition, $b_2$ is a two--form in Minkowski space, 
the 4D B--field. The gauge transformation of the two--form $b_2$ and the 
scalar $b_0$ are determined by the expansion of the three--form field 
strength~\cite{GrootNibbelink:2007ew}
\begin{equation}
H_3 = d b_2 + \Omega_{YM} - \Omega_L \,+\, \omega_2\, d B_0
\,+\, i\mathcal{F}  \big( d b_0 - 2 \tr[H_V iA]  \big),
\end{equation}
where $\Omega_{YM}$ and $\Omega_L$ are the Yang--Mills and Lorentz
Chern--Simons three forms, respectively. This implies that $B_0$ has
no  anomalous variations, only $b_2$ and $b_0$ can take part in the 4D 
Green--Schwarz mechanism. In particular, $b_0$ transforms under a gauge 
transformation as 
\begin{equation}
\delta_\Lambda b_0 = -2 \tr[H_V \Lambda], 
\label{AxionVariation}
\end{equation}
where $\Lambda$ is the gauge parameter. Therefore, the scalar $b_0$ 
and the Poincar\'e--dual of $b_2$  can be interpreted as axions, since 
the anomaly cancellation on the resolution occurs via the usual coupling 
of the B--field \cite{Green:1984sg}
\begin{equation}
\frac{1}{(2\pi i)^3}\hspace{-12pt}
\int\limits_{\mathbbm{M}^4\times\mathcal{M}^3} \hspace{-12pt}
B_2 X_8 \supset  
\frac{1}{(2\pi i)^3}\hspace{-12pt}
\int\limits_{\mathbbm{M}^4\times\mathcal{M}^3} \hspace{-12pt}
\left( b_2 X_{6,2} + i \mathcal{F} b_0 X_{4,4}\right) = 
\int\limits_{\mathbbm{M}^4}\left( b_2 X^{uni}_2 + b_0 X^{non}_4\right) \;.
\end{equation}
The dual of $b_2$ is the model--independent axion $a^{mi}$, 
because its existence does not depend on the particular internal manifold. 
The scalar  $b_0$ defines  the model--dependent axion 
$a^{md}$, i.e.\ $a^{md} = b_0$. The model--dependent axion is a 
localized state, as the field strength $i \cF$ becomes strongly peaked 
at the singularity in the orbifold limit. This means that it should
be interpreted as a twisted state from the orbifold perspective.

\subsubsection*{Relations between the various axions}

\fgRelationAxions

The matching of the spectra involves field redefinitions using a superfield 
$T$ associated to the blow--up mode. We showed that $T$ transforms 
with a shift under $\text{U}(1)$--gauge transformations. This means
that the imaginary part of $T$ transforms like an axion, which we
denote as $a^T$. We now investigate whether $a^T$ can be interpreted
as the model--dependent axion $a^{md}$ of the corresponding resolution
and to what extend the heterotic axion $a^{het}$ is related to the
model--independent axion $a^{mi}$. A schematic picture of the context
in which these different axions are defined is given in figure~\ref{fig:3}.

The field redefinitions are necessary to obtain the matching of the orbifold and 
blow--up spectra implying a modification of the anomaly polynomial for the 
heterotic orbifold model. First of all, we have to take into account that 
the orbifold gauge group is broken in the blow--up. Thus, the 
anomaly polynomial should also be re--expressed in terms of the new unbroken 
gauge group factors. The anomaly polynomial 
$\hat I^{het}_6 = X^{het}_2\cdot~X_{0,4}$ describes the anomaly of the
heterotic orbifold before the field redefinition but after the branching.

Moreover, the field redefinitions generate a new anomaly polynomial 
$\hat I^{red}_6 = i q F X^{red}_4$. Since they modify the U(1) charges 
of twisted superfields $\hat I^{red}_6$ is 
proportional to the field strength $i F$ of this U(1). Thus, the anomaly 
polynomial of the blow--up equals the sum of $\hat I^{het}_6$ and the
contribution from the anomalous field redefinition:
\begin{equation}\label{eq:anomalyrel}
\hat I^{het}_6 + \hat I^{red}_6 = \hat I^{blow}_6 = \hat I^{uni}_6 +
\hat I^{non}_6.
\end{equation}
In Table~\ref{tab:anomal} we list $\hat I^{het}_6$, model by model, 
computed from the orbifold model spectra, and $ \hat I^{uni}_6$ and 
$\hat I^{non}_6$, computed as discussed above.

The anomaly cancellation on the orbifold after the field redefinitions 
involves the heterotic axion $a^{het}$ and the localized twisted axion $a^{T}$.
Equation~(\ref{eq:anomalyrel}) implies the relation between the couplings of 
the various axions
\begin{equation}\label{eq:axionrel}
a^{het}X_{0,4} + a^T X^{red}_4 = a^{mi} X_{0,4} + a^{md} X^{non}_4\;.
\end{equation}
For a given model all the four--forms $X$ can be computed and 
(\ref{eq:axionrel}) yields a system of linear equations for each 
group factor. This system can be solved and results in the following 
relations between the various axions
\begin{equation}\label{eq:axions}
a^{mi} = a^{het} + \alpha\, a^{T}\;,\quad a^{md} = \beta\, a^{T}\;,
\end{equation}
with $\alpha,\beta$ in general being model dependent constants. 
The normalization of the axions is chosen such that for all 
$\mathbbm{C}^3/\mathbbm Z_3$ models $\beta=-\frac{1}{16}$ by
requiring that the blow--up modes always carry the same charge. Thus,
only the coefficient $\alpha$ is  model dependent and listed in the
last column  of Table~\ref{tab:anomal}.

\tbAnomPolyZthree

\subsection{D--terms in directions of anomalous U(1)'s}
\label{sec:Dflatness}

Since there is always a single twisted chiral superfield getting a vev, 
there can be a non--vanishing D--term only for one broken gauge symmetry 
generator. Moreover, since such a field is just a singlet of the non--Abelian
blow--up gauge group, there is a D--term only for a combination of the 
U(1)'s under which such a singlet $\Psi_q$ is charged. The presence
of such a  D--term is consistent: the non--vanishing D--term on
the blown--up orbifold corresponds to an FI--term on the resolution. 
In spite of the original orbifold having at most a single anomalous U(1) and 
thus a single FI-term, the resolved models can have two. 
The second one is just the counterpart of the D--term generated by the vev.
Hence, we conclude that D--flatness is guaranteed for all generators except 
the one corresponding to the broken U(1). But this non--vanishing D--term 
is required to make the FI--terms coincide: on the level of local blow--ups, 
we match two dynamically unstable models.

Let us comment on how it is possible that a
configuration chosen to be supersymmetric, i.e.\ which satisfies
the Hermitian Yang--Mills equations~\eqref{eq:YM}, leads to
non--vanishing D--terms. As was emphasized in~\cite{Blumenhagen:2005ga},
the Hermitian Yang--Mills equations get loop corrections precisely
when anomalous U(1)'s are present on smooth compactifications. In the
analysis of this paper we have ignored such loop effects in
the blow--up. The presence of non--vanishing D--terms for anomalous U(1)'s is
simply signaling this.

We will see at the end of the next section,
Section~\ref{sec:DflatnessExamples}, that D--flatness can be ensured 
in the compact case. There, we will use the local models (with $D\neq
0$) as building blocks for the construction of compact ones and
present various methods to obtain D--flatness afterwards.

\section{Blowing up the compact  $\boldsymbol{T^6/\mathbbm{Z}_3}$ orbifold}
\label{sec:compactorbifold}

The local study of orbifold singularities captures a lot of the physics
of compact orbifolds. The compact case has some important new
aspects as we demonstrate by studying the blow--up of the
$T^6/\mathbbm{Z}_3$ orbifold. The latter is a space which is flat
everywhere except at the 27 fixed points. For later use we enumerate 
the fixed points as $f = (f_1, f_2, f_3)$ with $f_i = 0,1,2$. The fixed 
point $0=(0,0,0)$ is obviously localized at the origin. The index $i$ labels 
the three complex $T^2$ directions. The fixed points are singular and 
the singularity is identical to the $\mathbbm C^3/\mathbbm{Z}_3$
singularity studied in the previous section. Thus, a sensible resolution 
of $T^6/\mathbbm{Z}_3$ can be constructed by cutting an open patch
around each singularity and replacing it with the smooth space studied
above.

To perform this procedure in detail one has to face the following
complicating issues: first of all one has to worry whether the gluing
process can be carried out properly. Constructing the blow--up of
$T^6/\mathbbm{Z}_3$ by naively joining 27 resolutions of
$\mathbbm{C}^3/\mathbbm Z_3$ with finite volume seems to lead to a
space that is not completely smooth. We ignore this complication by
assuming that a more complicated smooth gluing procedure 
exists, and that for essentially topological questions (e.g. what models
do exist and what are their spectra?) this procedure can be trusted. 
As we are not only gluing together the 
$\mathbbm C^3/\mathbbm Z_3$ blow--ups but also the bundles on
them, we have to confirm that the resulting bundle on the resolution
of $T^6/\mathbbm Z_3$ actually exists. There are two different ways of
analyzing this: we can check various consistency conditions ensuring 
the existence or, from the orbifold point of view, we have to
show that F-- and D--flat directions are allowed by the (super)potential
of the compact orbifold theory.

To systematically investigate these issues, we first show that
resolutions of compact orbifold models without Wilson lines are
possible. Next, we review properties of $\mathbbm Z_3$ orbifold models
with Wilson lines and their resolutions. We finish this section by two 
examples: the first example considers the blow--up of an orbifold with a 
single Wilson line, illustrating the gluing procedure of the gauge bundle. 
The second one examines an orbifold with two Wilson lines and defines 
an MSSM--like model. Therefore, it is phenomenologically interesting to 
see whether this model can exist in the blow--up.

\subsection{Resolution of the $\boldsymbol{T^6/\mathbbm{Z}_3}$
orbifolds without Wilson lines} 
\label{sec:withoutWilson}

To obtain the smooth resolution of an orbifold without Wilson lines, 
the first possibility is to choose the same U(1) bundle embedding 
at each fixed point. In such a case, the local consistency conditions 
are enough to guarantee the existence of the bundle.
Indeed, the only extra conditions on the bundle would come from the
Bianchi identity integrated on the new compact 4--cycles, which are 
generated by the gluing and thus ``inherited'' from $T^6$.
On the other hand, these new 4--cycles are obtained by combining the
non--compact 4--cycles of the resolved $\mathbbm C^3/\mathbbm Z_3$ 
singularities. However, for this resolution 
(see~\cite{Nibbelink:2007rd,Nibbelink:2007pn}), 
the local Bianchi identity on $\mathbbm C^3/\mathbbm Z_3$ 
implies the Bianchi identity on these non--compact 4--cycles. Thus,
the local consistency conditions ensure that the new consistency
conditions, due to the gluing, are satisfied. 
Therefore, all local models can be naturally extended to global ones, 
with spectra given by 27 copies of the local spectra. 
On the orbifold, this resolution is characterized by requiring that 
identical twisted states at all fixed points acquire non--vanishing vevs 
of the same magnitude and identical orientation.

From the orbifold perspective, it requires a little more work to show
that this blow--up exists. D--flatness does not constitute a problem: 
the auxiliary field $D^a$ is simply the sum of the local fixed
point contributions $D^{(f)a}$. Since at all fixed points identical
twisted states, the blow--up modes, attain exactly the same vev, the
individual D--terms $D^{(f)a}$ are all the same. For the compact models 
investigated here all D--terms vanish, except possibly the ones 
associated with the local anomalous U(1)'s, analogously to the
non--compact models studied before. For the anomalous
U(1)'s the same comment holds as for the non--compact situation, see 
subsection~\ref{sec:Dflatness}.

F--flatness of the compact blow--up does not automatically follow from
F--flatness of the local $\mathbbm{C}^3/\mathbbm{Z}_3$ blow--ups, because the
superpotential of the compact orbifold is much richer than its
non--compact counterpart. Of course, all local fixed point couplings that
were allowed on $\mathbbm C^3/\mathbbm Z_3$ are still allowed. But
since the $R$--symmetry group is reduced in the transition from the
non--compact to the compact orbifold as U(3)$_R \rightarrow \mathbbm{Z}_3^3$,
new local interactions at a single fixed point can appear. Moreover,
there is the possibility of non---local interactions involving twisted
states living at different fixed points.

Most arguments in subsection~\ref{sec:Fflatness} were based on the
existence of certain gauge invariant operators and therefore do still
apply in the compact case. For example, the blow--up A$\rightarrow$AI 
exists because~\eqref{E6invariant} yields vanishing F--terms for all fields 
if only the singlet gets a vev. In the compact case we have to take
non--local interactions into account,
\begin{equation}\label{eq:branching27ABC}
\mathcal W \sim 
\sum_{f,g,h}
\boldsymbol{27}^{(f)}\times
\boldsymbol{27}^{(g)}\times
\boldsymbol{27}^{(h)}
\sim  \sum_{f,g,h} 
\boldsymbol{16}^{(f)}_{-1}\times
\boldsymbol{16}^{(g)}_{-1}\times
\boldsymbol{10}^{(h)}_2 + 
\boldsymbol{10}^{(f)}_2\times 
\boldsymbol{10}^{(g)}_2\times 
\boldsymbol{1}^{(h)}_{-4}\;,
\end{equation}
where the sum over the different fixed points $f,\,g,\,h$ is
restricted by the space group selection
rule~\cite{Hamidi:1986vh,Kobayashi:1990mc}. Because
at all fixed points only the singlets $\boldsymbol{1}^{(f)}_{-4}$ get
vevs, all F--terms still vanish. Hence, we only have to worry about
gauge invariant superpotential terms that do not have an analog on the
non--compact orbifold.

The only case where new (and relevant) interactions arise on the compact 
orbifold, which did not exist in the non--compact version, is the standard
embedding AS. Because of the reduction of the $R$--symmetry group to 
$\mathbbm{Z}_3^3$ there is now a cubic gauge invariant term in the superpotential
\begin{equation}
\mathcal W \sim \sum_{\alpha, f,g,h} \epsilon^{klm} 
G^{(f)\alpha}{}_{k}G^{(g)\alpha}{}_{l}G^{(h)\alpha}{}_{m} + \ldots \;.
\end{equation} 
As argued in~\cite{Hamidi:1986vh}, this superpotential allows 
the same F--flat vev as in the non--compact case: $G = v \Identity$.

This analysis shows that a simultaneous blow--up of all 27 fixed
points, where the same blow--up mode at each fixed point acquires the
same non--vanishing vev, allows for D-- and F--flatness. It is therefore
possible -- and  straightforward -- to construct consistent resolutions
of compact orbifolds from the resolutions of the local ones, which
were studied in the previous sections.

\subsection{Orbifolds with Wilson lines}
\label{sec:withWilson}

Even though the description of orbifold models with Wilson lines is
well--known \cite{Ibanez:1986tp}, we give here a detailed review to be
able to emphasize similarities as well as differences compared to the 
description of blow--ups in the next subsection.

In compact orbifold models with multiple singularities, there can be 
different gauge embedding shifts $V_{orb}^{(f)}$ at each
fixed point $f$. Each of these shifts satisfies the local version of the modular
invariance requirement  
\begin{equation}\label{modinv}
(V_{orb}^{(f)})^2=0\,\text{mod}\,6.
\end{equation}
This means that, in the case of $T^6/\mathbbm{Z}_3$, the model is locally completely 
determined by the gauge groups and spectra listed in Table~\ref{tab:match}.

The possibility of having different local gauge shifts can also be encoded in
the language of discrete Wilson lines defined as 
$A_{orb}^{(fg)}=V_{orb}^{(g)}-V_{orb}^{(f)}$ among two fixed points
$f$ and $g$. However, not all local shifts $V_{orb}^{(f)}$ are
independent due to geometrical constraints.  As is
well--known~\cite{Gmeiner:2002es}, any local shift $V_{orb}^{(f)}$ can
be represented as  $V_{orb}^{(f)} \equiv V_{orb} + f_i A^{(i)}_{orb}$, 
where we define a global orbifold shift $V_{orb}=V_{orb}^{(0)}$ and
the three discrete Wilson lines
$A^{(i)}_{orb}=V_{orb}^{(i)}-V_{orb}^{(0)}$ of the $\mathbbm{Z}_3$
orbifold in the three complex directions.\footnote{With slight stretch
of notation we use $i=(\delta_{1i},\delta_{2i},\delta_{3i})$ to
indicate the fixed point which lies in the $i$th complex 
$T^2$.} The $\equiv$ symbol means that the two sides
of the equation are equal up to $3\Lambda$, where $\Lambda$ is a
generic element of the root lattice of $\text{E}_8\times\text{E}'_8$.
These vectors satisfy $\mathbbm{Z}_3$ periodicities 
\begin{equation}
3 V_{orb} \equiv 3 A^{(i)}_{orb} \equiv 0 
\end{equation} 
and the rewritten modular invariance conditions 
\begin{equation} 
(V_{orb})^2 = 0\,\text{mod}\,6\;, \qquad 
(A^{(i)}_{orb})^2 = 0\,\text{mod}\,6\;,\qquad 
2 V_{orb}A^{(i)}_{orb} = 0\,\text{mod}\,6\;. 
\end{equation}
Sitting at a fixed point $(f_1,f_2)$ of the first two
tori but freely moving in the third one these conditions imply
\begin{equation}
\label{geocondition}
V_{orb}^{(f_1,f_2, 0)}+V_{orb}^{(f_1,f_2,1)}+V_{orb}^{(f_1,f_2,2)}\equiv 0\;.
\end{equation}
Similar conditions have to be imposed for the other choices of tori.

At each fixed point the local action $V_{orb}^{(f)}$ generates a
(different) gauge symmetry breaking.  The resulting 4D gauge group is
the one surviving all local projections simultaneously. Using the splitting of the
generators of $\text{E}_8\times\text{E}'_8$ into Cartan elements $H^I$
and other elements $E^p$, where $p$ denotes the 16--dimensional root vector of  
$E^p$, such that $[H^I, E_p]=p^I E_p$, the effective 4D gauge group is 
determined by $V_{orb}^{(f)}\cdot p = 0\,\text{mod}\,3$ for each fixed
point $f$.  These conditions can be rewritten in terms of the gauge
shift and Wilson lines as 
\begin{equation}
\label{orbproj}
V_{orb}\cdot p = 0\,\text{mod}\,3,\,\,\,\,\,\,
A^{(i)}\cdot p = 0\,\text{mod}\,3, \,\,\,\,\text{for}\,\,
i=1,\,2,\,3\;, 
\end{equation}
and provide an efficient way of characterizing the effective 4D
gauge group.

We have to distinguish between localized and delocalized matter when
describing the spectrum on $T^6/\mathbbm{Z}_3$.  The twisted
states localized in the fixed points are organized into representations
of the larger gauge group at the respective fixed point, determined by
$V_{orb}^{(f)}$ {\em only}. They are listed in Table~\ref{tab:match}. 
Since $T^6/\mathbbm{Z}_3$ has no fixed planes
or lines, only untwisted matter is delocalized.\footnote{This is not
generically true; most orbifolds have sectors of delocalized twisted
matter, e.g.\ the second twisted sectors in some $T^6/\mathbbm{Z}_{2n}$
orbifolds.} It feels the action of {\it all} local projections.

\subsection{The resolution of $\boldsymbol{T^6/\mathbbm{Z}_3}$ models with 
Wilson lines}

In this section we describe how to construct smooth resolutions of
compact orbifold models in the presence of discrete Wilson
lines. After summarizing the basic matching principle, we study the
consistency conditions that must be enforced due to the global
properties of the compact space. Finally, we explain how to
compute the spectrum of the resolved models.

\begin{figure}[t]
\begin{center}
\raisebox{-4ex}{\scalebox{.5}{\mbox{\input{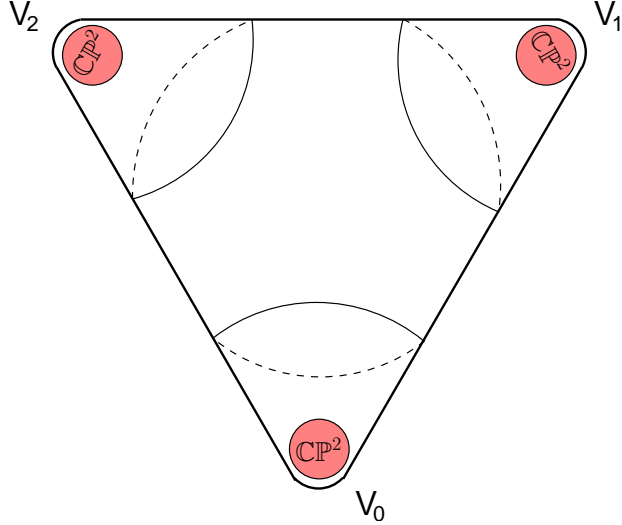}}}} 
\end{center}
\caption{\label{loops}
A schematic two dimensional cross section of the resolved
$T^6/\mathbbm Z_3$ orbifold is depicted. The fixed points are replaced
by smooth surfaces that contain $\mathbbm{CP}^2$'s. 
}
\end{figure}

Having discrete Wilson lines on an orbifold essentially corresponds to
wrapping different local fluxes on the $\CP^2$'s inside the resolved
space, i.e. choosing different embedding vectors $V^{(f)}$ at
different resolved singularities. A schematic picture of the resolved
situation is depicted in Fig.~\ref{loops}.

Constraints on the possible fluxes come from the local Bianchi
identities, related to the localized 4--cycles corresponding
to the exceptional divisors:  at each fixed point $f$
we have a condition\footnote{As explained in the previous section,
the new conditions due to the presence of new compact 4--cycles are
automatically satisfied once the local conditions are.}
\begin{equation}
(V^{(f)})^2=12.
\end{equation}
Moreover, new conditions are due to the fact that the gauge bundles are not
localized, but rather extend over the whole space. Hence, the gluing
of different patches requires the various gauge backgrounds on 
non--trivial overlaps to be related in a consistent way. Therefore, we
consider open patches $U^{(f)}$ and $U^{(g)}$ around the resolutions of
orbifold singularities labeled by $f$ and $g$ with gauge
configurations $\fA_1^{(f)}$ and $\fA_1^{(g)}$, respectively. The transition
function $g^{(fg)} = (g^{(gf}){}^{-1}$ describes the relation between
the two gauge one--form potentials on the intersection of the two patches:  
\begin{equation}
\fA_1^{(g)} ~=~ g^{(gf)} (\fA_1^{(f)} + \text{d} )g^{(fg)}~.
\end{equation}
Given a point where (any) three patches $f$, $g$ and $h$ overlap, we need
$g^{(fg)} g^{(gh)} g^{(hf)} =1$.
Moreover, we can identify the transition function $g^{(fg)}$ in the case of 
a $\text{U}(1)$ gauge bundle with a function
$A^{(fg)}$ between the two fixed points $f$ and $g$ as
\begin{equation}
g^{(fg)} ~=~ e^{2\pi i\, A^{(fg)I} H_I/3}~.
\end{equation}
The function $A^{(fg)}$ is generically not constant. However, in the 
blow--down limit it becomes constant and can be identified with a
discrete Wilson line $A^{(fg)}_{orb}$ on the orbifold between 
the fixed points $f$ and $g$. In this limit, we have 
$A^{(fg)} \equiv A^{(fg)}_{orb}$ and for that reason we may refer to the
function $A^{(fg)}$ as a Wilson line on the resolved space.

The co--cycle condition $g^{(fg)} g^{(gh)} g^{(hf)} =1$ can be expressed in
terms of the Wilson lines as 
\begin{equation}
        A^{(fg)}+A^{(gh)}+A^{(hf)}\equiv 0\;.
\end{equation}
This condition applies to any manifold. It states conditions for the existence 
of a flux in the case a space cannot be covered with a single open patch.  

The construction of such Abelian gauge bundles on the resolution obtained 
from gluing the local patches, as discussed above, leads to a more general 
class of models than those
that are obtained as blow--ups of global orbifold models: the gauge bundle
on the resolution of  $T^6/\mathbbm Z_3$ descents 
down to the orbifold gauge bundle only if an identification
\begin{equation}
\label{matchVcmpct} 
V^f \equiv V^f_{orb} 
\end{equation} 
can be made at each of the resolved fixed points of
$T^6/\mathbbm Z_3$, as explained in~\eqref{matchV}. Thus, 
the geometric condition~\eqref{geocondition} has
to be imposed on the resolution shifts, too, and we have
\begin{equation}
\label{res-iden}
V^{(f_1,f_2,0)}+V^{(f_1,f_2,1)}+V^{(f_1,f_2,2)}\equiv 0 
\end{equation}
and corresponding expressions after permutations of the tori. 
Resolution models, that do not satisfy this condition, nevertheless
define valid supergravity compactifications, even though they can not be
associated with a global orbifold model. This shows that such
global orbifold constructions may lead only to a restrictive class of
models. Since the motivation of this paper is to study the blow--up of
such global orbifolds, we enforce the conditions~\eqref{res-iden}. 

Given the consistency conditions on the bundles, we can study how to
compute the spectra in the 4D models. Gauge bosons of the global
unbroken 4D gauge group are distributed over the whole resolution manifold. 
Let us consider the blow--up of two singularities $f$ and $g$, each surrounded
by an open patch. Assume that the patches have a non--vanishing
overlap. Again, we consider gauge configurations $\fA_1^{(f)}$ and
$\fA_1^{(g)}$ on the patches $U^{(f)}$ and $U^{(g)}$, respectively. 
Since we are interested in the resulting zero modes, we can assume
that the non--trivial topology is encoded in the background gauge
configurations $\cA_1^{(f)}$ and $\cA_1^{(g)}$ only, i.e.\ they satisfy
the same relation as above
\begin{equation}
\cA_1^{(g)} ~=~ g^{(gf)} (\cA_1^{(f)} + \text{d} )g^{(fg)}~.
\end{equation}
The full gauge configurations $\fA_1^{(f)}$ is written as a sum of the
background $\cA_1^{(f)}$ plus perturbations $A_1^{(f)}$: 
\begin{equation}
\fA_1^{(f)} ~=~ \cA_1^{(f)} + A_1^{(f)}~. 
\end{equation}
In this expansion we only take the untwisted modes into account.
This means that we find 
\begin{equation}
A_1^{(g)} ~=~ g^{(gf)} A_1^{(f)} g^{(fg)}
\quad \Rightarrow \quad 
\left\{ 
\begin{array}{l}
A_1^{(g)I} ~=~ A_1^{(f)I}~, 
\\[2ex] 
A_1^{(g)p} ~=~ e^{2\pi i\, A^I_{fg} p_I} \, A_1^{(f)p}~, 
\end{array}
\right. 
\end{equation}
after expanding the perturbations as $A_1^{(f)} = A_1^{(f)I} H_I + A_1^{(f)p} E_p$, 
with the notation of the $H_I$ and $E_p$ explained above~(\ref{orbproj}). If we
assume the overlap region of the two patches $U^{(f)}$ and $U^{(g)}$ to be
far away from the blown up singularities, the zero modes of the
perturbations are essentially constant modes. Because the constant zero
modes on both sides of the gluing region $U^{(f)} \cap U^{(g)}$ can be
connected and stay a zero mode, they simply have to be equal. This means
that the phase must be trivial: $A^{(fg)} \cdot p =0$. 
Hence, in terms of bundle shift $V$ and Wilson lines $A^{(f)} = A^{(f0)}$ we
find the projection conditions
\begin{equation}
\label{rescond}
V\cdot p =0,\,\,\,\,\,A^{(f)}\cdot p =0,\,\,\,\text{for}\,\,f \neq 0.
\end{equation}
As compared to the maximally four projection conditions for the
effective 4D gauge group on the orbifold, we see that there are generically
more and stronger conditions on the surviving 4D gauge group on the
resolution.

The main reason for the additional gauge symmetry breaking on the
resolution is that the conditions~(\ref{rescond}) are not ``$\text{mod}\,\,3$", 
as they were in the orbifold case. This means that we cannot neglect 
(triple multiples of) $\text{E}_8\times\text{E}_8$ lattice vectors 
and reduce to four projections at most. In particular, this implies that an
orbifold {\em irrelevant Wilson line}, i.e.\ just being three times an
$\text{E}_8\times\text{E}_8$ lattice vector, can have a non--trivial
effect on the resolution gauge group. In this case the same U(1) bundle 
is chosen at each fixed point, but they are differently aligned in the 
$\text{E}_8 \times \text{E}'_8$. From the orbifold point of view 
this choice corresponds to identical twisted states at all fixed points acquiring 
non--vanishing vevs of the same magnitude, but different orientation. 
It will also be shown later that this can help to ensure D--flatness for all
U(1)'s.

As an example of this situation we can consider the gauge embeddings 
$V=(-2^3,0^5)(0^8)$ and $V'=(1^3,3,0^4)(0^8)$. They are both 
consistent and give rise to the same resolution model labeled as AI with 
gauge group $\text{SO(10)}\times \text{U(3)}\times \text{E}_8$. It is the 
resolution of the orbifold model A with gauge group
$\text{E}_6\times\text{SU(3)}\times\text{E}_8$. 
Nevertheless, if we consider a compact model with one resolved singularity
equipped with a $V$-- and another one with a $V'$--embedding,
such that the trivial Wilson line $A=(3^4,0^4)(0^8)$ relates them, then
the resulting gauge group is not $\text{SO(10)}\times \text{U(3)}\times \text{E}_8$,
but rather $\text{SO(8)}\times\text{U(1)}\times \text{U(3)}\times \text{E}_8$.
This can also be confirmed from the orbifold perspective, when the $\rep{27}$--plets of
these two fixed points develop vevs for different components.

Finally, we describe the consequence of this for the matter on the
blow--up of $T^6/\mathbbm Z_3$ with Wilson lines.  Locally, the
delocalized matter was identified by the fact that it has a fractional
multiplicity factor, $\frac{1}{9}$ (or multiples), see
section~\ref{sec:localmatching}. Because it is distributed over all
patches, it feels projection conditions due to the transition
functions between the patches.  
Thus, given a resolved singularity, say 0, we have to impose 
\begin{equation}
\label{eq:matterprojections}
A^{(f)}\cdot p =0 \text{ mod }3,\,\,\,\text{for}\,\,f \neq 0
\end{equation}
on its
delocalized matter. The localized matter, with integral multiplicity,
does not reach the overlap regions with the other patches and
therefore feels no further projection conditions. Hence, the matter
representations of the localized matter just branch with respect to
the global unbroken 4D gauge group.

\subsection{One Wilson line model with three anomalous U(1)'s}
\label{sec:wilsonex}

\fgCompactBlowupVEVs

In the following we give a specific example of an orbifold model in
the presence of a discrete Wilson line, and study one of its blown up
versions. On the resolution the model has three anomalous 
U(1)'s. The bulk universal and the local model--dependent axions 
are all involved in the anomaly cancellation.

To make the general discussion more explicit, we consider the
model obtained from the $T^6/\mathbbm{Z}_3$ orbifold with gauge shift
$V_{orb}=(2,2,0^6)(2,0^7)$ and one Wilson line
$A_{orb}=(0,-4,2,0^5)(-2,0^7)$ in the first complex
torus.~\footnote{For recent work about the computation of orbifold
spectra with Wilson lines see for example
\cite{Forste:2004ie,Ploger:2007iq}.} First we look at the orbifold 
and then investigate its resolution. Due to the Wilson line on the
orbifold, the $27$ fixed points are grouped together in three sets of
nine fixed points each. The three sets are characterized by the local
shift vectors $V_{orb}$, $V_{orb}+A_{orb}$ and $V_{orb}+2A_{orb}$
respectively. The same local gauge group and charged matter is present
at all nine fixed points of each set. Details are given in
Table~\ref{tab:table2}, where representatives of the three sets of
fixed points are identified by their space group representatives $g_1,
g_2$ and $g_3$, respectively.

The next task is to find a resolution model that, in the blow down
limit, reduces to this orbifold model. We find that at the $g_1$
singularities, we have to choose the CI resolution, with gauge bundle
defined by the blow--up shift $V_1 = V_{orb}$; at the $g_2$
singularities the AI resolution, with $V_2=V_{orb}+A_{orb}$. Finally, 
at the $g_3$ singularities we have to choose again resolution CI,
but with a different shift $V_3 = V_{orb} + 2A_{orb} + 3\Lambda$,
where $3\Lambda=(0,6,-6,0^5)(0^8)$ represents, from the orbifold
perspective, an irrelevant Wilson line, which is nevertheless crucial 
to ensure that $V_3$ satisfies the local Bianchi identity. This 
``irrelevant'' Wilson line leads to additional gauge symmetry breaking 
on the resolution. The local gauge group and the chiral matter on each 
of the three sets of nine patches can be found in table~\ref{tab:match}. 
The different bundle vectors $V_1$, $V_2$ and $V_3$ combined lead to
further symmetry breaking of the local gauge groups at the 27 
resolved fixed points to the global 4D gauge group: 
\begin{equation}
\text{SO(10)}\times \text{SO(14)}\times \text{U(1)}^4\;.
\label{G_ex1}
\end{equation}
Consequently, the representations of the local spectrum on each of the
different fixed point resolutions becomes 
\begin{equation}
\begin{array}{ll}
g_1:~ CI: ~ & \frac{1}{9}\left[
(\boldsymbol{16};\boldsymbol{1})_{(2,2,0,\text{-}1)} + 
(\boldsymbol{10};\boldsymbol{1})_{(2,2,0,2)} + 
(\boldsymbol{1};\boldsymbol{1})_{(2,2,0,\text{-}4)} \right] +
 3\,(\boldsymbol{1};\boldsymbol{1})_{(4,2,\text{-}2,0)}\;,
 \\[2ex] 
g_2:~ AI:~ & \frac{1}{9}\left[
(\boldsymbol{16};\boldsymbol{1})_{(2,2,0,\text{-}1)} + 
(\boldsymbol{10};\boldsymbol{1})_{(2,2,0,2)} + 
(\boldsymbol{1};\boldsymbol{1})_{(2,2,0,\text{-}4)} \right] + 
(\boldsymbol{16};\boldsymbol{1})_{(0,0,0,3)} 
\\[2ex]
& + \, 
3\left[ \,(\boldsymbol{1};\boldsymbol{1})_{(\text{-}2,\text{-}2,0,4)} + 
(\boldsymbol{1};\boldsymbol{1})_{(0,2,2,4)} + 
(\boldsymbol{1};\boldsymbol{1})_{(2,0,\text{-}2,4)} \right]\;,
\\[2ex]
g_3:~ CI:~ & \frac{1}{9}\left[
(\boldsymbol{16};\boldsymbol{1})_{(2,2,0,\text{-}1)} + 
(\boldsymbol{10};\boldsymbol{1})_{(2,2,0,2)} + 
(\boldsymbol{1};\boldsymbol{1})_{(2,2,0,\text{-}4)} \right] +
3\,(\boldsymbol{1};\boldsymbol{1})_{(2,4,2,0)}\;.
\end{array} 
\label{localResSpec}
\end{equation}
Comparing this with Table~\ref{tab:match}, the localized states (with
integral multiplicities) are simply branched to representations of the unbroken 4D gauge
group, while some delocalized states (with multiplicity 1/9) are
projected out. Because these delocalized states live everywhere on the
compact resolution, their spectra at the three types of patches are
all the same. The complete resolution spectrum is obtained by
multiplying each line of~\eqref{localResSpec} by nine.

We can also study this resolved model from the orbifold blow--up
perspective: we select a single twisted field per fixed point that
attains a vev chosen along a F--flat direction, but some D--terms are
induced in order to match the FI--terms of the resolution.\footnote{
For complete F-- and D--flatness, we can choose another vacuum 
configuration, defined by the monomial $(\boldsymbol{27}, \boldsymbol{1})_{(2,2,0)}^2 
(\boldsymbol{1}, \boldsymbol{1})_{(-4,0,0)} (\boldsymbol{27}, \boldsymbol{1})_{(0,0,0)} 
(\boldsymbol{1}, \boldsymbol{1})_{(0,-4,0)}$. This means that the additional
untwisted field  $(\boldsymbol{27}, \boldsymbol{1})_{(2,2,0)}$ gets  
a vev leading to a further gauge symmetry break down.} 
We determine the gauge symmetry breaking induced by
this. Each set of singularities $g_i$ has a different blow--up
mode and gauge symmetry breaking:  
\begin{equation} 
\begin{array}{lrccl}
g_1:~ & \langle (\boldsymbol{1};\boldsymbol{1})_{(-4,0,0)}\rangle \neq
0\;:
~ & \text{E}_7\times \text{SO(14)}\times \text{U(1)}^2 & 
\rightarrow & \text{E}_7\times \text{SO(14)}\times \text{U(1)}\;, 
\\[2ex]  
g_2:~ & \langle (\boldsymbol{27};\boldsymbol{1})_{(0,0,0)} \rangle
\neq 0\;:~   
& \text{E}_6\times \text{SU(3)} \times \text{E}_8 & 
\rightarrow & \text{SO(10)}\times \text{U(3)}\times \text{E}_8\;, 
\\[2ex]  
g_3:~ & \langle (\boldsymbol{1};
\boldsymbol{1})_{(0,-4,0)} \rangle \neq 0\;:~  
& \text{E}_7\times \text{SO(14)}\times \text{U(1)}^2 & \rightarrow & 
\text{E}_7\times \text{SO(14)}\times \text{U(1)}\;. 
\end{array}
\end{equation} 
The global 4D gauge group can be obtained as the intersection of the
three local ones, and coincides with the one given in \eqref{G_ex1}.
By performing the appropriate field redefinitions on the orbifold,
given in Table~\ref{tab:table2}, the blown--up orbifold and the smooth
resolution model match perfectly.

Let us finally comment on the issue of anomalous U(1)'s of
this orbifold model in blow--up. As one can see from Table~\ref{tab:table2}, 
at each fixed point the blow--up mode induces a localized axion. We refer to
these axion superfields as $T_1, T_2$ and $T_3$, depending on which set 
of nine fixed points they belong to. Together with $b_2$, there can in
principle be four independent types of axions in the resolution model;
this theory could maximally accommodate four anomalous U(1)'s. Because
the anomaly polynomial  
\begin{gather}
\hat I_6^{res} ~=~  
216\, F_1^{an}\left(\frac{3}{2}(F_1^{an})^2+(F_2^{an})^2
+\frac{1}{2}(F^{no})^2 +\frac{1}{24}F_{10}^2-\frac{1}{32}\tr R^2\right)
\nonumber \\[2ex] 
~+~ 216\, F_2^{an}\left(\frac{3}{2}(F_1^{an})^2+7(F_2^{an})^2
+\frac{1}{2}(F^{no})^2
+\frac{1}{8}F_{10}^2-\frac{7}{96}\tr R^2\right) \;,
\end{gather}
with $F^{no}=F_1-F_2-2F_3$, is a sum of two factorized pieces, we
could infer that there are only two anomalous U(1)'s,
$F_1^{an}=F_1+F_2$ and $F_2^{an}=F_4$. (The corresponding 
charges  are defined in the caption of Table~\ref{tab:table2}.) However, if we
more physically define the number of anomalous U(1)'s as the number of
independent massive U(1) gauge fields, the number is three: three different  
vevs $v_1$, $v_2$ and $v_3$ break the U(1) symmetries $Q_1$, $Q_4$ 
and $Q_2$, respectively. The three axions $T_1, T_2$ and $T_3$ 
that do transform under three different combinations of the U(1)'s
couple to the corresponding gauge field strengths, leading to 
three massive gauge fields. We can confirm
this statement directly on the resolution by considering the gauge
transformations  
\begin{equation}
\delta_\Lambda b_0^{g_1} = -2 \tr[ Q_1 \Lambda]~,
\qquad
\delta_\Lambda b_0^{g_2} = -2 \tr[ Q_4 \Lambda]~,
\qquad
\delta_\Lambda b_0^{g_3} = -2 \tr[ Q_2 \Lambda]~,
\end{equation}
obtained from~\eqref{AxionVariation} for the local expansions of $B_2$
at the  resolutions of the different fixed points. Hence, these states
can be identified as 
\begin{equation}
b_0^{g_i} = 2\, T_i,  
\end{equation}
with the local axions $T_i$ from the orbifold blow--up.

\subsection{Can we blow--up a ${\mathbbm Z}_3$ MSSM model?}

We consider the ${\mathbbm Z}_3$ orbifold model with two Wilson lines
initially introduced in~\cite{Ibanez:1987sn}. This model is
interesting because it was one of the first string models with 
Standard Model gauge group and three generations of quarks and
leptons. A potential problem of this model is the set 
of vector--like exotics in the spectrum. Only if these exotic states can 
all be made heavy, the effective low energy spectrum will be 
identical to that of the MSSM. The way this may happen is by turning 
on appropriate vevs. As vevs of twisted states lead to blow--ups of 
the singularities on which they are localized, it is interesting to 
investigate blow--up versions of this model. Therefore, we assume that 
the blow--up of this model is generated by single vevs of twisted states 
at each of the 27 fixed points. This assumption guarantees that we can 
rely on the Abelian bundles, constructed in section \ref{sec:classification}, 
only. We focus on the question whether crucial properties of 
the MSSM are maintained in blow--up.

The work of~\cite{Ibanez:1987sn,Casas:1988se} revealed the presence of two 
hypercharge candidates amongst the eight U(1) factors of the model and an 
resulting ambiguitiy of identifying the MSSM particle spectrum. However, for either choice the orbifold theory cannot be 
completely blown up without breaking hypercharge. To
resolve all singularities simultaneously, one
blow--up mode has to be chosen per fixed point. Table~1 of~\cite{Casas:1988se} implies
that all the states at the fixed point $(n_1,n_3) = (-1,-1)$ carry the
same charge under both hypercharge candidates. Hence, by blowing up
this singularity, we inevitably break hypercharge. There is only one 
way to avoid the end of any phenomenology in this
orbifold model in full blow--up: the Higgs doublet $H_1$ of the
MSSM at $(-1,-1)$ has to obtain a vev. Hence, the blow--up procedure has the
interpretation of electroweak symmetry breaking. As far as we have
been able to confirm, such a scenario still does not lead to a
phenomenologically acceptable situation, because the vanishing of
all the D--terms requires the 
vev of $H_1$ to be of the order of the compactification scale,
i.e.\ far too large.

For this reason we explore a second possibility and resolve all
singularities except the one at $(n_1,n_3) = (-1,-1)$. This partial
resolution can be performed in an entirely  F-- and D--flat way, in all
U(1) directions including the anomalous one and without breaking the
hypercharge. For F--flatness, we need higher orders in the superpotential to
guarantee that the derivative of the superpotential has a zero. 
For concreteness, consider the situation in which the fields listed in
Table~\ref{tab:Z3-2WLblowupmodes} all have non--vanishing vevs.  
Their gauge invariant monomial 
\begin{equation}
h_2\, (h_{10})^2\, (h_{14})^2\, h_{15}\, (h_{17})^3\, h_{21}\, (h_{23})^3\, (h_{24})^2
\end{equation}
corresponds to the following relation between the vevs~\cite{Buccella:1982nx} 
\begin{equation}
\sqrt{6} h_2 = \sqrt{3}h_{10} = \sqrt{3}h_{14} = \sqrt{6}h_{15} = \sqrt{2}h_{17} = \sqrt{6} h_{21} = \sqrt{2} h_{23} = \sqrt{3}h_{24}\;,
\end{equation}
which ensures D--flatness. In this configuration, the hypercharge is identified to be 
$Y =\frac 16 \big( \frac{1}{3} Q_1 - \frac{1}{2}Q_2 - Q_3 + Q_4 \big)$, 
so that none of the blow--up modes is charged under it.
Since $H_1$ is massless but does not constitute
a flat direction of the effective scalar potential away from this point
(i.e.\ at least as long as supersymmetry is not broken),
the Higgs cannot acquire a vev. Consequently, electroweak symmetry breaking
can only occur at low energies. Furthermore, in this vev configuration all extra 
$\text{U}(1)$'s are broken and all extra colour triplets acquire high masses from 
trilinear couplings. However, some of the other vector--like exotics stay massless 
at this order in the superpotential.
Thus finally, neither the singular orbifold
nor the everywhere smooth resolution of all the fixed points, but the partial
blow-up to this \textit{hybrid model} can potentially save phenomenology.

\tbVevsZthreeMSSM

\subsection{F-- and D--terms for compact blow--ups}
\label{sec:DflatnessExamples}

We have mainly focused on compact resolutions with multiple 
anomalous U(1)'s and corresponding FI--terms. From the orbifold 
perspective, we have seen that these terms can be interpreted as 
non--vanishing D--terms induced by vevs of the blow--up modes. This
situation is exactly the same as explained in
section~\ref{sec:Dflatness}. In the following, we will discuss various
possibilities to obtain  stable resolutions by finding 
orbifold blow--ups corresponding to vacua with $F=D=0$.

The first method was discussed in the previous section, where it was necessary to blow--up the orbifold only 
partially in order to obtain $F=D=0$. This may seem a rather easy way
out. A more interesting possibility is that some additional matter
fields, either twisted or untwisted, take non--vanishing vevs. When more than one twisted state develops
vevs at a single fixed point, we expect a non--Abelian gauge
background to be generated on the resolution, as discussed at the end
of section~\ref{sec:Fflatness}. A vev for an untwisted state leads to
a {\em continuous} Wilson line. An example of the latter case was
presented in Section~\ref{sec:wilsonex}, where the vev of the untwisted
state $(\boldsymbol{27}, \boldsymbol{1})_{(2,2,0)}$ yielded a stable 
vacuum.

The general idea of a third method is to perform different blow--ups 
of degenerate fixed--points, i.e. of fixed points not distinguished 
by Wilson lines from the orbifold perspective. This can be achieved by
choosing different blow--up modes at the various fixed--points. They may be
either contained in different types of non--Abelian representations or
in the same ones, but in different components. This allows for choosing 
the vevs at the different fixed points such that all D--terms vanish globally.

We can exemplify the latter possibility by considering the 
blow--up of the compact orbifold B without Wilson lines, see 
Section \ref{sec:withoutWilson}. Here, the blow--up mode is 
contained in the representation $(\rep{1},\rep{3};\rep{1},\rep{3})$, 
denoted by the matrix $C$. D--flatness can be guaranteed by 
assigning a vev of the same magnitude, but {\it different orientation} to each of the fields $C_i$, localized at one 
of the 27 fixed points $i=1,\ldots,27$. 
This corresponds to a gauge invariant monomial of the form
\begin{equation}
\prod_{i=1}^{27} C_i\,,
\end{equation}
breaking the $\text{SU}(3)^2$ factors of the 4D orbifold gauge 
group to $\text{U}(1)^4$.  Furthermore, F--flatness $F=0$ can 
be achieved at isolated points using higher order couplings in 
the superpotential yielding stable SUSY preserving vacua.

\section{The resolution of the $\boldsymbol{\mathbbm{C}^2/\mathbbm Z_2}$ orbifold}
\label{app:C2overZ2}

Orbifold singularities of the form $\mathbbm{C}^n/\mathbbm Z_n$ are 
resolved by a generalization of the procedure given in the previous section. 
In this way, it is possible to approach the resolution of the 
$\mathbbm{C}^2/\mathbbm Z_2$ 
singularity that is phenomenologically relevant given that many appealing
4D orbifold models are based on compactifications on orbifolds having
$\mathbbm{C}^2/\mathbbm Z_2$ subsectors.

In \cite{Nibbelink:2007rd} the explicit form of the resolution curvature and
bundles were given for the $\mathbbm{C}^2/\mathbbm Z_2$ singularity, as well as a study
of the matching of 4D models arising from the SO(32) heterotic string quantized on
the orbifold and on the resolution. We do not give the details of that derivation,
rather, we summarize the relevant results in Table~\ref{tab:matchC2}.
In the table we give the gauge group and spectra of the three orbifold models
2A, 2B and 2C in the first column. In the second column we list those of the 
three models 2AI, 2BI and 2CI obtained by compactifying 10D SO(32) 
supergravity on the resolution. Again, the multiplicities are
fractional and multiples of $1/16$ for untwisted (non--localized) states,
and are integer or half integer for twisted (localized) states.
As in the $\mathbbm{C}^3/\mathbbm{Z}_3$ case, there is no direct matching 
of the spectra. They should be compared only after the blow--up mode has 
developed a vev. This vev induces a Higgs mechanism on the orbifold side of 
the matching partially breaking the gauge symmetry. In the mechanism, parts 
of the Higgs field are ``eaten'' by the gauge bosons becoming massive.
From the resolution perspective, this can be seen in the multiplicities $7/8$
for the states corresponding to the Higgs fields. Indeed, such a multiplicity
should be understood as an integer number (the old twisted field multiplicity)
reduced by $1/8$, since $1/8$ of the twisted field is incorporated in the
massive gauge fields\footnote{Note that the gauge bosons are delocalized, 
thus their multiplicity is $1/16$ times a factor of $2$, since they from doublets 
of the internal $\text{SU}(2)$ holonomy.}.
All the other states match after the field redefinition given in the third column
of Table~\ref{tab:matchC2}. No extra state becomes massive due to the fixed
chirality of the hyper multiplets in 6D.

The analysis of the matching at the pure spectrum level is thus not different 
from the $\mathbbm{C}^3/\mathbbm{Z}_3$ case. On the other hand, the 
study of the flatness of the blow--up mode, as well as that of anomaly 
cancellation, is technically very different. The first issue is due to the structural 
difference between 4D and 6D SUSY, the second one due to the fact that 
anomaly cancellation in 6D may proceed via two different diagrams, giving rise 
to different mechanisms: one is mediated by scalars (or four--forms in a dual 
picture), the other by two--forms. In the first situation the anomalous gauge 
boson gets a mass, in the second case it does not. We approach these aspects 
in the forthcoming sections.

\tbMatchingBlowupZtwo

\subsection{Flatness of the zero mode}

The flatness study of the blow--up mode in the 6D case is different
from the 4D case due to the difference in the structure of the scalar
potentials. In the 4D case the potential for the scalars in the chiral
multiplets is derived from the gauge interactions (D--term potential)
and from the superpotential (F--term potential).
In the 6D case the whole potential for the scalars in the
hyper multiplets is encoded in the gauge interactions.
Indeed, the scalars $\Phi_i$ can be organized into doublets of a
global SU(2) symmetry. The D--terms are defined as
\begin{equation}
D^{a,\rho} = \sum \Phi^*_{i,M} \sigma^\rho_{MN} t^{a,ij} \Phi_{j,N}\;,
\end{equation}
where $\sigma^\rho$ are the three Pauli matrices related to the global
SU(2) and $t^a$ denotes a generator of the gauge interactions.
Then, the scalar potential is just $V=D^2$.

A detailed study of such a potential was given in \cite{Honecker:2006qz}.
There it was shown that the D--term related to the U(1) symmetry,
under which the blow--up mode is charged, cannot be zero in case
a single blow--up mode is introduced. On the contrary, as argued
in~\cite{Honecker:2006qz}, flatness is always ensured in case more
than a single mode is switched on, but not at the same fixed point,
i.e.\ in case we have a mutual blow--up of more than one singularity.

\subsection{Anomaly cancellation}

Let's begin by studying the anomaly cancellation for the resolved
models by integrating the 10D anomaly polynomial over the resolution. 
The resulting 6D anomaly polynomial is
\begin{equation}
 \hat I_8 = X_{0,4}\cdot X^{uni}_4 + X^{non}_2 \cdot X^{non}_6 
\equiv  \hat I^{uni}_{8} + \hat I^{non}_8 \;,
\end{equation}
where 
\begin{eqnarray*}
X_{0,4} &=&  \text{tr}R^2 - \text{tr}(i F)^2\;,\\
X^{uni}_4 &=& -\frac{1}{96}\left[\text{Tr}\left[(\frac{1}{8}H^2_V - \frac{3}{40})(i F)^2 \right] - 
\frac{1}{4\cdot(30)^2}\left( \text{Tr}(H_V i F)\right)^2  - \frac{3}{8}\text{tr}R^2\right]\;, \\
X^{non}_2 &=& -2\text{tr}\left[ H_V i F\right]\;, \\
X^{non}_6 &=& -\frac{1}{192}\left[ \frac{1}{6}\text{Tr}(H_V (i F)^3) - 
\frac{1}{120}\text{Tr}(H_V i F)\left[\frac{1}{15}\text{Tr}(i F)^2 + \text{tr}R^2 \right]\right] \;.  
\end{eqnarray*}
The generic polynomials given above are computed for the three
$\mathbbm{C}^2/\mathbbm Z_2$ models, see Table~\ref{tab:anomalC2}. 
The structure of the two terms is different, indicating different diagrams as 
source for the anomaly, and slightly different Green--Schwarz mechanisms. 
Indeed, an anomaly term factorized as $X_2\times X_6$ is canceled either 
by a scalar axion having an anomalous variation, or, in the dual picture, by 
a four--form axion. In both cases, this extra degree of freedom can
be reabsorbed, fixing the gauge, into the longitudinal component of a 
massive vector boson, in a way similar to the Higgs mechanism or to the 
standard 4D Green--Schwarz mechanism. Instead, an anomaly term 
factorized as $X_4\times X_4$ is canceled by a two--form axion.

For the heterotic orbifold models the anomaly polynomial is always of the form
$X_4\times X_4$ (see Table~\ref{tab:anomalC2} for the explicit form for the three
$\mathbbm{C}^2/\mathbbm Z_2$ models), and the Green--Schwarz mechanism involves a
single ``axion'', i.e. the 6D components of the untwisted $B$--field.
So, as in the 4D case, a matching between the anomaly cancellation mechanisms 
requires to take into account the field redefinitions that have to be performed during
the blowing--up procedure. Again, a relation between the anomalies on
the orbifold and on the resolution has to hold,
\begin{equation} \label{I_8match}
\hat I^{het}_8 + \hat I^{red}_8 = \hat I^{blow}_8 = \hat I^{uni}_8 + \hat I^{non}_8\;.
\end{equation}
One observes that $\hat I^{red}_8$ is factorized as $X_2 \times X_6$. Thus, the
corresponding anomaly cancellation mechanism induces a mass for the
anomalous U(1). This is in agreement with the fact that, from the orbifold
perspective, the blow--up corresponds to a Higgs mechanism giving a mass
to the broken U(1).

\tbAnomPolyZtwo

About the matching of the axions and anti--symmetric tensor fields, 
things are less straightforward, since
in 6D the dual of $b^{blow}_2$ cannot be interpreted as a massless scalar.
Indeed, the matching leads to a relation among the B--fields
$b^{blow}_2\equiv b_2$ and $b_2^{orb}\equiv h^{het}_2$, 
and the axions $a^{md}$ and $a^T$, which reads as
\begin{equation}
b^{blow}_2 X_{0,4} + a^{md} X^{non}_6 = b^{orb}_2 X_{0,4} + a^T X^{red}_6\;.
\end{equation}
As explained, the axions $a^T$,  $a^{\text{md}}$ and the $B$--fields $h_2^{het}$,
$b_2$ are forms of different degree,
namely $0$--forms and two--forms. Hence, one would expect relations only between forms
of the same degree, i.e. $a^T\sim a^{md}$ and $h_2^{het}\sim b_2$. This would
require also $X_4^{uni}\sim X_4^{het}$ and $X_6^{red}\sim X_6^{non}$, 
what is in general not true, see table \ref{tab:anomalC2}. Only model 2BI fulfills
this condition, and in such a case we deduce
\begin{equation} \label{axionrelations2B}
b_2=h^{het}_2,\,\,\, a^{md}=-8a^T.
\end{equation}
For models 2AI and 2CI, instead, both $X_4^{uni},\, X_4^{\text{het}}$
and  $X_6^{red},\,X_6^{non}$ are not proportional, and more work is needed
in order to relate the axions and $B$--fields. We address this issue in the following paragraphs.

Let us begin the discussion by listing the gauge transformations of the axions and
$B$--fields 
\begin{eqnarray} \label{gaugetrafo}
\text{Orbifold}&:& 
\delta_{\Lambda}h^{het}_2\nearrow X_{0,4},\,\,\,\,\, \delta_{\Lambda}a^T\nearrow 3(iF);\\ 
\text{Resolution}&:& \delta_{\Lambda}b_2\nearrow X_{0,4},\,\,\,\, 
\delta_{\Lambda}a^{md}\nearrow X_{2}^{non},
\end{eqnarray}
where the arrow $\nearrow$ indicates that the descent equations have to be
used, and $\delta_{\Lambda}$ denotes a combined gauge and Lorentz transformation.
The gauge transformation of the $B$--fields is model--independent and fixed, on both 
the orbifold and the resolution side. Hence, we have to change our description of the
physical degrees of freedom to overcome this rigidity of~(\ref{gaugetrafo}) to be
able to formulate relations between the fields on both sides.
This can be achieved by passing to the dual forms, denoted by the
two--forms $\tilde{h}^{het}_2$, 
$\tilde{b}_2$ and the four--forms 
$\tilde{a}_4:=\tilde{a}^T_4$, $\tilde{c}_4:=\tilde{a}^{\text{md}}_4$.
The duality transformation interchanges couplings and gauge transformations
(on the level of the anomaly polynomial). Thus, the gauge transformations read as 
\begin{eqnarray} \label{gaugetrafosdual}
\text{Orbifold}&:& \delta_{\Lambda}\tilde{h}^{het}_2\nearrow X_{4}^{\text{het}},\,\,\,\,\, 
\delta_{\Lambda}\tilde{a}_4\nearrow X_6^{\text{red}};\\
\text{Resolution}&:& \delta_{\Lambda}\tilde{b}_2\nearrow X_{4}^{\text{uni}},\,\,\,\, 
\delta_{\Lambda}\tilde{c}_4\nearrow X_{6}^{\text{non}},
\end{eqnarray}
whereas the couplings are given by the forms in (\ref{gaugetrafo}).
Precisely these gauge transformations will help us to relate the orbifold and 
resolution fields appropriately. We have to express
$\tilde{h}^{het}_2$ and $\tilde{a}_4$ by their counterparts
$\tilde{b}_2$, $\tilde{c}_4$ and the Abelian gauge fields, such that
both sides of these expressions transform identically under gauge
transformations. Hence, their gauge transformations produce the same
term on the level of the anomaly polynomial and the anomaly
(\ref{I_8match}) is canceled.

We now apply this method to model 2CI to deduce the axion and $B$--field 
relations. The gauge transformations for $\tilde{b}_2$ and
$\tilde{h}^{het}_2$ are given in this case, such that the difference is just
\begin{equation}
\delta_{\Lambda}\left(\tilde{b}_2-\tilde{h}^{het}_2\right)\nearrow 
X_4^{\text{uni}}-X_4^{\text{het}}=-\frac{15}{4}(iF)(iF'),
\end{equation}
compare Table~\ref{tab:anomalC2}.
Hence we can apply the descent equations to obtain the relation between the
$B$--fields
\begin{equation} \label{b_2h_2}
\tilde{b}_2=\tilde{h}^{het}_2-\frac{15}{4}(iA)(iA').
\end{equation}
To relate also $\tilde{c}_4$ and $\tilde{a}_4$ we make the ansatz
\begin{equation} \label{ansatz}
\tilde{c}_4=-\frac{1}{8}\tilde{a}_4+\gamma (iF')\tilde{h}^{het}_2+(iA')Y_3
\end{equation}
for a free parameter $\gamma$ and a three--form $Y_3$. 
Using~(\ref{gaugetrafosdual}) we can consider the terms induced by a gauge 
transformation on the level of the anomaly polynomial. This yields a factorized 
expression, which justifies our ansatz~(\ref{ansatz}) 
\begin{eqnarray}
X_6^{\text{non}}+\frac{1}{8}X_6^{\text{red}}&=&(iF')
\left[-\frac{5}{16}\tr_{\bf{15}}\left(iF_{\bf{15}}\right)^2-\frac{15}{8}(iF)^2-
\frac{25}{2}(iF')^2+\frac{5}{32}\tr R^2\right]\\
&\stackrel{!}{=}&\gamma (iF')\tilde{h}^{het}_2+(iF')Y_4,
\end{eqnarray}
where the explicit expressions for $X_6^{\text{non}}$ and $X_6^{\text{red}}$
and $Y_4:=dY_3$ were used. 
However, naive use of the descent equations for $(iF')^3$ yields 
$(iA')\left[(iA')(iF')\right]=0$, because $iA'$ is an Abelian gauge field. 
Thus, this term  can not be obtained from the anomalous variation of
$(iA')Y_3$ as $Y_3$ had to contain $\left[(iA')(iF')\right]$.
Consequently, $(iF')^3$ has to be generated by the anomalous transformation
of $\tilde{h}^{het}_2$ that is determined by $X_4^{\text{het}}$.
We can use this observation to determine $\gamma$. Fortunately, 
$X_4^{\text{het}}$  contains a term $\frac{25}{8}(iF')^2$, see 
table~\ref{tab:anomalC2}, such that we can write
\begin{eqnarray}
&X_6^{\text{non}}+\frac{1}{8}X_6^{\text{red}}
\supset -\frac{25}{2}(iF')^3\stackrel{!}{=}\gamma\frac{25}{8}(iF')^3
\subset \gamma(iF')X_4^{\text{het}}&\\
&\Longrightarrow \gamma=-4.& 
\end{eqnarray}
Now, we are able to calculate $Y_4$ by plugging in all results. Hence, we obtain
\begin{eqnarray}
(iF')Y_4&=&X_6^{\text{non}}+\frac{1}{8}X_6^{\text{red}}+4(iF')X_4^{\text{het}}\\
&=&(iF')\left[-\frac{5}{16}\tr_{\bf{15}}\left(iF^2_{\bf{15}}\right)-
\frac{3}{4}(iF)^2+\frac{15}{2}(iF)(iF')+\frac{9}{64}\tr R^2\right].
\end{eqnarray}
Finally, this results can be used for the descent equations to obtain the expression
for $Y_3$ in terms of the Chern--Simons forms for the various characteristic classes 
appearing in $Y_4$. This concludes the explicit matching of the axions and
$B$--fields in~(\ref{b_2h_2}) and~(\ref{ansatz}).

The methods presented above can also be used to relate the axions and $B$--fields
of model 2A and its resolution 2AI. However, the inconvenient
numerical effort of this calculations will prevent us from addressing this here.
Let us just note that analogous relations to the ones given above can
be deduced and that $\tilde{c}_4=-\frac{1}{8}\tilde{a}_4+\ldots$ occurs.   
Of course, these new descriptions reproduce the results given above for 
model 2BI as we can write $\tilde{h}^{het}_2=\tilde{b}_2$ and $\tilde{c}_4=-
\frac{1}{8}\tilde{a}_4$. They automatically have the same gauge
transformations. After  dualization of these relations we recover our
old result~(\ref{axionrelations2B}), because the factor $-\frac{1}{8}$
converts to $-8$. Hence, it seems also for the 6D models to be a
model--independent statement that $a^{\text{md}}=-8a^T+\ldots$,
although the precise relations are much more complicated.

\section{Conclusions and outlook}
\label{sc:concl}

We have compared heterotic string models on orbifolds with
supergravity models on smooth compact spaces obtained by
resolving the corresponding orbifold. Our motivation 
was to extend the physics of the orbifold constructions to regions in the
moduli space ``far'' away from the orbifold point. Our main focus was on 
heterotic  $\text{E}_8\times\text{E}_8'$ supergravity models assembled on 
resolutions of the $T^6/\mathbbm Z_3$ orbifold, 
allowing for Wilson lines to be present. To prepare for this analysis we 
considered models on resolutions of the non--compact orbifold
$\mathbbm C^3/\mathbbm  Z_3$ before turning to the compact case. 
We achieved full agreement between orbifold and
resolved models, at the level of gauge interactions, massless spectrum
and anomaly cancellation.

First, we reviewed the construction of the non--compact  
$\mathbbm C^3/\mathbbm Z_3$ resolution, equipped with a gauge flux, that we completely 
classified in case of an U(1) bundle background.  
Then, on the level of $\text{E}_8\times \text{E}_8'$ heterotic supergravity, we matched
each $\mathbbm C^3/\mathbbm Z_3$ resolution model, built by employing Abelian 
gauge bundle backgrounds, with an orbifold model. This extends the results of~\cite{Nibbelink:2007rd} 
and~\cite{GrootNibbelink:2007ew} in the $\text{SO}(32)$ 
context. We achieved the matching by single--vev blow--ups of the orbifold model, i.e. by giving a vev to a {\em single} 
twisted state along an F--flat direction of the potential.  
We emphasized that a single orbifold model has different blow--ups, if multiple twisted states 
are present.

After this we investigated the matching in detail. We demonstrated the 
fundamental importance of the blow--up mode to obtain full 
agreement of the spectra as well as the anomaly cancellation mechanisms in 
both theories. First, we used the blow--up mode to perform field redefinitions 
of the matter fields on the orbifold, so that the massless spectrum, including 
the U(1) charges, coincided with the one on the resolution.  Then, we showed 
how the Higgs mechanism caused by the vev of the blow--up mode relates 
the maximally two anomalous U(1)'s on the resolution to the potential 
anomalous U(1) of heterotic orbifolds. A detailed analysis of the identification 
of the axions entering anomaly cancellation was performed:
In the resolved models two axions take part in the cancellation of the two 
anomalous U(1)'s. Both are among the three zero--modes of the 10D 
anti--symmetric tensor field $B_{MN}$. Two zero--modes fill the whole 
internal space and are just constant in the blow--down limit. But only one of 
those has an anomalous variation and therefore corresponds to a 4D axion. 
From the orbifold perspective this {\it untwisted} state can be 
identified as the model--independent axion. The third zero--mode of
$B_{MN}$ is  peaked around the singularity, hence this
model--dependent axion should be identified with a {\it twisted}
state. It is precisely the counterpart of the blow--up mode on the orbifold. 
All this shows that the blow--up mode is the crucial ingredient in the
matching between the orbifold and resolution models.

Since we matched SUSY orbifold models with SUSY resolved models, it 
was decisive to ensure F-- and D--flatness of the blow--up mode. 
This could always be achieved up to a single D--term:  precisely the
one corresponding to the U(1) gauge symmetry that is anomalous 
on the resolution but not on the orbifold. This is to be expected, because 
when a U(1) anomaly is cancelled via the Green--Schwarz mechanism, 
a Fayet--Iliopoulos term is generated in the potential. 
The resolution model accommodates two anomalous U(1)'s and 
 two FI terms, while the orbifold model has only a single 
anomalous U(1) and a single FI term. From the orbifold perspective, the additional FI term on 
the resolution is induced by the vev of the blow--up mode.

Next we focused on the central theme of this work, the study of
resolution models of the compact $T^6/\mathbbm Z_3$ orbifold. 
We resolved it by replacing each of the 27 singularities
with a copy of the smooth space used in the non--compact case. 
In the transition to the compact case the superpotential is modified: 
new couplings arise among states localized in different fixed points,
as well as among states at the {\it same} fixed point. Nevertheless
switching on the same blow--up mode at all fixed points is still F--flat. 
This demonstrates that  $T^6/\mathbbm Z_3$ orbifold models
in the absence of Wilson line can be blown up by simply joining 27
identical copies of the $\mathbbm C^3/\mathbbm Z_3$ resolution.

Discrete Wilson lines on compact orbifolds constitute a crucial ingredient
of heterotic orbifold model building. Therefore, reproducing them on 
smooth spaces provides a definite step forward, towards the 
construction of realistic models in blow--up. On a compact resolution, 
discrete Wilson lines correspond to the possibility of having different
gauge fluxes wrapping different cycles of resolved singularities.
In the large volume limit, or, equivalently, in the blow--down limit,
they are identified with the transition functions connecting patches
surrounding different resolved singularities. We have demonstrated how to 
compute the resulting gauge group and matter spectrum in the presence of such 
transition functions; they generically lead to more gauge symmetry 
breaking than discrete Wilson lines on orbifolds. In particular, we 
can have transition functions that correspond to trivial discrete Wilson lines 
from the orbifold perspective, but that still induce some gauge symmetry 
breaking in the resolved models. Then, we investigated the structure of anomaly 
cancellation in the presence of discrete Wilson lines. Contrarily to the 
non--compact case, more than two anomalous U(1)'s may be present.

As an application of our general principles, we considered the
resolution of a semi--realistic $T^6/\mathbbm Z_3$ MSSM model 
studied in~\cite{Ibanez:1987sn}. We found that no complete blow--up is
possible using U(1) fluxes without breaking the hypercharge
of the model.

Finally, we also considered the $\mathbbm C^2/\mathbbm Z_2$ orbifold, in the
same spirit as in the $\mathbbm C^3/\mathbbm Z_3$ case. We reviewed the matching 
of the spectra and the computation of anomaly polynomials done 
in \cite{Honecker:2006qz}, showing how to match, via field
redefinitions, the blown--up and resolved spectra at the level of both
non--Abelian and Abelian charges, and thus how to match the anomaly
cancellation mechanism, completing the results of~\cite{Honecker:2006qz}. 
The corresponding identification of axions and anti--symmetric tensor fields 
is much more cumbersome than in the $\mathbbm C^3/\mathbbm Z_3$ 
case; we explained details of the matching in one specific example.

As mentioned above, our main motivation was to
extend the power of the orbifold construction to regions of the moduli
space where direct string quantization is very difficult to perform. 
This is crucial if we want to address issues like moduli stabilization
or the study of the ``landscape'' of heterotic models.
Moreover, this is essential when the orbifold model is driven away
from the orbifold point by a Fayet--Iliopoulos term corresponding to 
an anomalous U(1). Since many resolved models constructed in this work 
contain anomalous U(1)'s, they do not provide stable endpoints of 
such flows. As we indicated, stable vacua can be obtained by vevs 
at different fixed points conspiring to lead to vanishing D--terms, 
or by multiple vevs at some fixed points. We believe that stable
points in the moduli space with multiple vevs can be brought forth by smooth
non--Abelian flux compactifications. 

We found that many of the blown--up models can be reproduced as 
resolved models with U(1) fluxes. But there is also a large class of
orbifold blow--ups that do not lead to Abelian bundles. Of course, the
``standard embedding'' provides an example of a non--Abelian SU(3)
background. However, there should be many non--Abelian flux
models induced by giving vevs to more than one twisted states in the
blow--up procedure. It would be interesting to understand, what
background flux can be reproduced by such {\em multiple} vevs. 
We hope that the classification and matching we performed, 
will prove helpful to determine the topological properties
of the required non--Abelian bundles.

\subsection*{Acknowledgments}

We are grateful to Kang--Sin Choi, Arthur~Hebecker, Christoff~L\"udeling,
Hans~Peter~Nilles, Sa\'ul~Ramos--S\'anchez, Michael~Ratz and Michael~Schmidt 
for discussions and comments. This work was partially supported by the 
European Union 6th framework program MRTN-CT-2004-503069 ``Quest for 
unification'', MRTN-CT-2004-005104 ``ForcesUniverse'', MRTN-CT-2006-035863 ``UniverseNet'' 
and SFB-Transregio 33 ``The Dark Universe'' by Deutsche Forschungsgemeinschaft (DFG).
The work of MT is supported by the European Community through the contract
N~041273 (Marie Curie Intra-European Fellowships), he would like to thank the
Institute for Theoretical Physics of Heidelberg for hospitality during the completion
of the work.

\end{document}